\documentclass[acmsmall]{acmart}

\AtBeginDocument{%
  }

\usepackage{pifont}
\usepackage{bbding}
\usepackage{subfigure}
\usepackage{fontawesome}
\usepackage{adjustbox}
\usepackage{threeparttable}
\usepackage[ruled,linesnumbered]{algorithm2e}

\usepackage{longtable} 
\usepackage{booktabs}  

\usepackage{array}
\usepackage{listings}

\newcommand{\tech}{DESIL}
\newcommand{\optnum}{\textit{optnum\_each}}
\newcommand{\diffnum}{\textit{diffnum}}

\newcommand{\codeIn}[1]{{\ttfamily\tiny #1}}
\newcommand{\qc}[1]{{\color{olive}\ding{46}[Qingchao: #1]}}

\begin{document}


\title{\tech{}: Detecting Silent Bugs in MLIR Compiler Infrastructure}


\author{Chenyao Suo}
\affiliation{%
  \institution{Tianjin University}
  \city{Tianjin}
  \country{China}}
\email{chenyaosuo@tju.edu.cn}

\author{Jianrong Wang}
\affiliation{%
  \institution{Tianjin University}
  \city{Tianjin}
  \country{China}}
\email{wjr@tju.edu.cn}

\author{Yongjia Wang}
\affiliation{%
  \institution{Tianjin University}
  \city{Tianjin}
  \country{China}}
\email{yongjiawang@tju.edu.cn}

\author{Jiajun Jiang}
\affiliation{%
  \institution{Tianjin University}
  \city{Tianjin}
  \country{China}}
\email{jiangjiajun@tju.edu.cn}

\author{QingChao Shen}
\affiliation{%
  \institution{Tianjin University}
  \city{Tianjin}
  \country{China}}
\email{qingchao@tju.edu.cn}

\author{Junjie Chen}
\authornote{Junjie Chen is the corresponding author.}
\affiliation{%
  \institution{Tianjin University}
  \city{Tianjin}
  \country{China}}
\email{junjiechen@tju.edu.cn}


\begin{abstract}



MLIR (Multi-Level Intermediate Representation) compiler infrastructure provides an efficient framework for introducing a new abstraction level for programming languages and domain-specific languages. 
It has attracted widespread attention in recent years and has been applied in various domains, such as deep learning compiler construction.
Recently, several MLIR compiler fuzzing techniques, such as MLIRSmith and MLIRod, have been proposed. 
However, none of them can detect silent bugs, i.e., bugs that 
incorrectly optimize code silently.
The difficulty in detecting silent bugs arises from two main aspects:
(1) \textbf{UB-Free Program Generation}: Ensures the generated programs are free from undefined behaviors to suit the non-UB assumptions required by compiler optimizations.
(2) \textbf{Lowering Support}: Converts the given MLIR program into an executable form, 
enabling execution result comparisons, 
and selects a suitable lowering path for the program to reduce redundant lowering pass and improve the efficiency of fuzzing.
To address the above issues, 
we propose \tech{}. 
\tech{} enables silent bug detection by defining a set of UB-elimination rules based on the MLIR documentation 
and applying them to input programs to produce UB-free MLIR programs. 
To convert dialects in MLIR program into the executable form,  \tech{} designs a lowering path optimization strategy to convert the dialects in given MLIR program into executable form. 
Furthermore,
\tech{} incorporates the differential testing for silent bug detection. To achieve this, it introduces an operation-aware optimization recommendation strategy into the compilation process to generate diverse executable files.
We applied \tech{} to the latest revisions of the MLIR compiler infrastructure. It detected 23 silent bugs and 19 crash bugs, 
of which 12/14 have been confirmed or fixed.

\end{abstract}

\maketitle

\section{Introduction
}
\label{sec:intro}

The MLIR (Multi-Level Intermediate Representation) compiler infrastructure is a powerful and extensible framework designed to facilitate compiler construction across diverse domains, including machine learning, high-performance computing, and hardware accelerators~\cite{mlir}.
By providing a structured representation at multiple abstraction levels, MLIR enables efficient transformations, optimizations, and target-specific code generation, making it a cornerstone of modern compiler design. 
However, given its growing adoption in critical applications such as deep learning and hardware synthesis, ensuring the correctness of MLIR is paramount. 
Particularly, bugs in MLIR can propagate through the compilation pipeline, leading to incorrect program execution, degraded performance, or even security vulnerabilities~\cite{DBLP:conf/issta/Suo00JZW24, DBLP:conf/kbse/WangCXLWSZ23}. 
Therefore, rigorous testing techniques are essential, ensuring that MLIR remains a robust and trustworthy infrastructure for compiler development and optimization.

Due to the unique characteristics of MLIR (such as its use of dialects to manage multi-level IRs and its proprietary data structures and semantics), traditional compiler testing techniques are largely inapplicable.
Therefore, in recent years, some testing techniques tailored to the MLIR compiler infrastructure have been proposed~\cite{DBLP:conf/issta/Suo00JZW24, DBLP:conf/kbse/WangCXLWSZ23}.
For example, MLIRSmith~\cite{DBLP:conf/kbse/WangCXLWSZ23} generates MLIR programs based on its grammar for the testing purpose.
MLIRod~\cite{DBLP:conf/issta/Suo00JZW24} mutates existing MLIR programs for testing, guided by the diversity of operation dependencies within MLIR programs.
However, these techniques are limited to detecting crash bugs, failing to capture silent bugs (also known as wrong code bugs~\cite{DBLP:conf/pldi/YangCER11}), which generate incorrect executable code without triggering crashes. 
This limitation arises due to the challenging issue of undefined behavior (UB) — a scenario where program execution lacks a well-defined outcome due to violations of language specifications, leading to unpredictable execution results~\cite{DBLP:conf/pldi/YangCER11}. 
Note that silent compiler bugs pose a severe risk, as they can go unnoticed during compilation and cause erroneous behaviors at runtime, potentially leading to critical failures in real-world applications.

In the literature, eliminating UB has been recognized as an important yet challenging task~\cite{DBLP:conf/pldi/YangCER11, DBLP:journals/pacmpl/LivinskiiBR20, DBLP:journals/pacmpl/LecoeurMD23, DBLP:journals/pacmpl/LivinskiiBR23}.
This challenge arises from the diverse root causes of UB — such as memory safety violations, uninitialized variables, integer overflows, and type mismatches — which can emerge at any stage of compilation and propagate silently through optimizations.
The unique characteristics of MLIR further exacerbate this problem.
Specifically, MLIR supports multiple dialects, each with its own operations, attributes, and verification rules, significantly expanding the scope of potential UB. 
Moreover, MLIR introduces dialect-specific UB root causes, such as shape inconsistency in \textit{memref} and \textit{linalg} dialects, which require specialized runtime checks and analysis for effective detection and elimination.
Unlike traditional programming languages, MLIR lacks dedicated UB detection tools, making even well-known UB issues more difficult to identify and mitigate.

Assuming UB-free MLIR programs can be obtained, using them to detect silent bugs still faces the compilation challenge — the process of transforming an MLIR program into an executable form (i.e., solely represented by the \textit{llvm} or \textit{spirv} dialect). 
This challenge arises because MLIR programs often require multiple lowering stages across different dialects (especially their operations) before reaching a fully executable representation. 
Specifically, an MLIR program may contain operations from various dialects, each necessitating specific lowering passes to transition into an executable representation. 
For ease of presentation, we call a sequence of lowering passes to transform a dialect operation to the specified executable dialect \textit{an operation-specific lowering path}, and a sequence of lowering passes to transform an MLIR program to the executable form \textit{a lowering path}.
Furthermore, new dialects can be introduced dynamically during the lowering process, leading to an expansive and evolving space of possible lowering paths.
While an exhaustive enumeration of all possible lowering sequences could theoretically ensure successful compilation, it would impose a significant efficiency bottleneck — a crucial factor in compiler testing~\cite{DBLP:conf/icse/ChenHHXZ0X16}.
Therefore, determining an appropriate lowering path is essential to balance compilation feasibility and testing efficiency, enabling more effective detection of silent bugs.

To bridge the gap in detecting silent MLIR bugs, we propose \textbf{\tech{}} (\textbf{DE}tecting \textbf{SIL}ent bugs), a novel technique that jointly generates UB-free programs and determines an optimal lowering path for each program to facilitate effective bug detection.
Specifically, to tackle the first challenge, \tech{} designs a set of MLIR program transformation rules to eliminate UB in UB-prone operations.
For example, to address the UB-prone operation of copying a {\tt memref} variable to another when both have dynamic shapes, \tech{} introduces a transformation rule that replaces the source {\tt memref} value with one that deterministically matches the destination variable's shape.
Notably, \tech{} tackles the UB issue through post-processing of already-generated MLIR programs, making it orthogonal to existing MLIR testing techniques. 
This allows \tech{} to be seamlessly combined with them, enhancing the detection of silent bugs and demonstrating its practicality.
To tackle the second challenge, \tech{} determines an optimal lowering path that prevents redundant or circular application of lowering passes, ensuring efficient compilation to an executable representation. 
Specifically, \tech{} first builds a mapping between lowering passes and dialect operations based on MLIR documentation, recording an operation-specific lowering path for each operation. 
Then, given an MLIR program, \tech{} determines its optimal lowering path by performing \textit{topological sorting} on the lowering passes derived from the operation-specific lowering paths of the program’s operations.

With these UB-elimination rules and the lowering-path optimization algorithm, \tech{} can effectively and efficiently compile an UB-free MLIR program into an executable form.
However, it is hard to directly determine whether the executable program is as expected. 
Therefore, to make \tech{} self-contained, we incorporate the differential testing mechanism into \tech{}.
Specifically, \tech{} introduces operation-aware optimization recommendation, which specifies optimization passes according to the opeartions in the given MLIR program, and obtains a set of executable programs produced by different optimization passes for differential testing.
Any inconsistent result produced by their executions are regarded as a silent bug detected by \tech{}.

To evaluate the effectiveness of \tech{}, we applied \tech{} to test the latest versions (from {\tt adbf21} to {\tt b6d5fa}) of the MLIR compiler infrastructure over approximately four months.
Specifically, we integrated \tech{} with MLIRSmith and MLIRod to process their generated MLIR programs, naming them \tech{}$_{\textit{smith}}$ and \tech{}$_{\textit{od}}$, respectively.
In total, \tech{} detected 42 previously unknown bugs, including 23 silent bugs and 19 crash bugs, of which 18 have been fixed and 26 confirmed by developers.
We further compared \tech{} with two enhanced state-of-the-art techniques, MLIRSmith$_{\textit{enhanced}}$ and MLIRod$_{\textit{enhanced}}$ (since neither of them can transition MLIR programs into executable forms to detect silent bugs), through five repeated 12-hour fuzzing sessions.
The results show that \tech{}$_{\textit{smith}}$ and \tech{}$_{\textit{od}}$ detected 29 and 38 bugs, respectively, outperforming MLIRSmith$_{\textit{enhanced}}$ (20) and MLIRod$_{\textit{enhanced}}$ (25), while also significantly reducing false positives in silent bug detection.
The latter techniques suffered from extremely high false positive rates (97.33\% and 96.96\%) due to the UB issue.
Additionally, our ablation study confirmed the essential contributions and practicality of \tech{}'s lowering path optimization and operation-aware optimization recommendation strategies.
For evaluating lowering path optimization strategy, 
we replaced this strategy with a random lowering pass selection strategy and designed two variants: \tech{}$_{\textit{smith}}^{\textit{w/o lower}}$ and \tech{}$_{\textit{od}}^{\textit{w/o lower}}$.
Through five repeated 12-hour fuzzing sessions, 
neither \tech{}$_{\textit{smith}}^{\textit{w/o lower}}$ nor \tech{}$_{\textit{od}}^{\textit{w/o lower}}$ successfully lowered any MLIR program within 50 lowering passes.
In contrast, \tech{}$_{\textit{smith}}$ and \tech{}$_{\textit{od}}$ required only 21 lowering passes on average to lower an MLIR program.
These results demonstrate the effectiveness of the lowering path optimization strategy.
For evaluating operation-aware optimization recommendation strategy, 
we replaced it with a random optimization selection strategy and designed two variants: \tech{}$_{\textit{smith}}^{\textit{w/o opt}}$ and \tech{}$_{\textit{od}}^{\textit{w/o opt}}$. 
We conducted five repeated 12-hour fuzzing sessions.
The results show that \tech{}$_{\textit{smith}}$ and \tech{}$_{\textit{od}}$ detected 29 and 38 bugs, respectively, 
significantly outperforming \tech{}$_{\textit{smith}}^{\textit{w/o opt}}$ (21) and \tech{}$_{\textit{od}}^{\textit{w/o opt}}$ (31).
These findings demonstrate the effectiveness of the operation-aware optimization recommendation strategy.

In this paper, we makes the following main contributions:
\begin{itemize}
    \item We propose \tech{}, the first testing technique designed to detect silent bugs in the MLIR compiler infrastructure.

    \item We design a set of MLIR program transformation rules to eliminate UB-prone operations in any given MLIR program, enabling the feasible generation of UB-free MLIR programs for effective silent bug detection.

    \item We introduce a lowering-path optimization strategy by performing topological sorting on the lowering passes associated with the operations in a given MLIR program, identified through MLIR documentation analysis.
    This strategy ensures efficient compilation into an executable form, optimizing the lowering process by eliminating redundant lowering.

    \item We evaluate \tech{} on the latest versions of the MLIR compiler infrastructure, uncovering 42 previously unknown bugs, of which 18/26 have been fixed/confirmed by developers. 
    Notably, we have publicly released our experimental data and implementation on our project homepage~\cite{DESILrepository}.
    
\end{itemize}

\section{Background and Motivation}

\begin{figure}[t]
  \centering
  \subfigure[An example of MLIR program]{
    \label{fig:moti1}
    \includegraphics[width=0.9\textwidth]{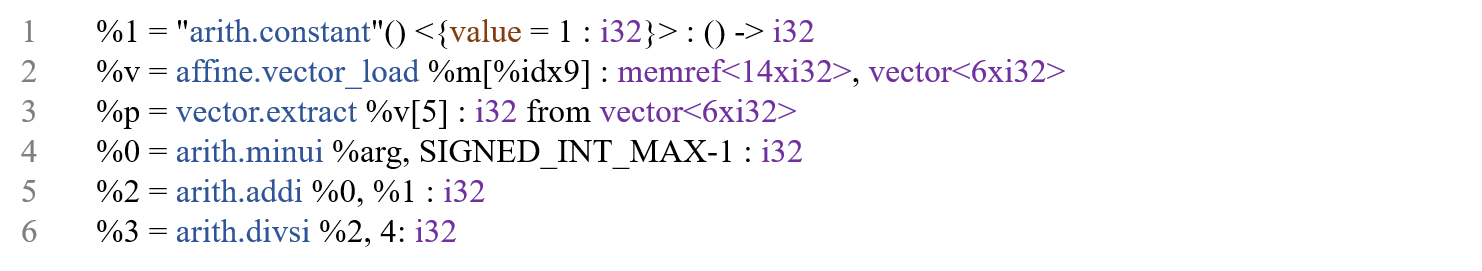}
  }

  \subfigure[Lowered MLIR program (for lines 4-6 in Figure~\ref{fig:moti1}) by applying ``-convert-arith-to-llvm'']{
    \label{fig:moti2}
    \includegraphics[width=0.9\textwidth]{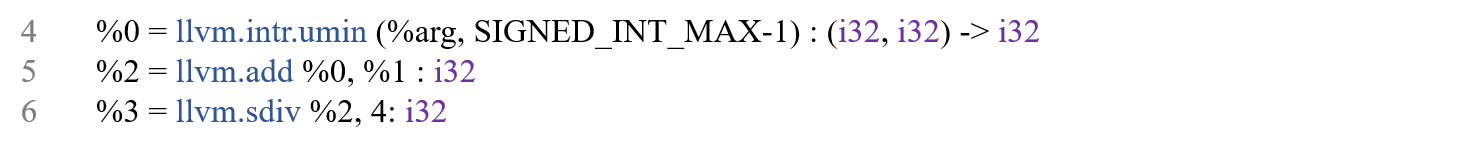}
  }

    \subfigure[Optimized MLIR program (for lines 4-6 in Figure~\ref{fig:moti1}) by applying ``-arith-unsigned-when-equivalent'']{
    \label{fig:moti3}
    \includegraphics[width=0.9\textwidth]{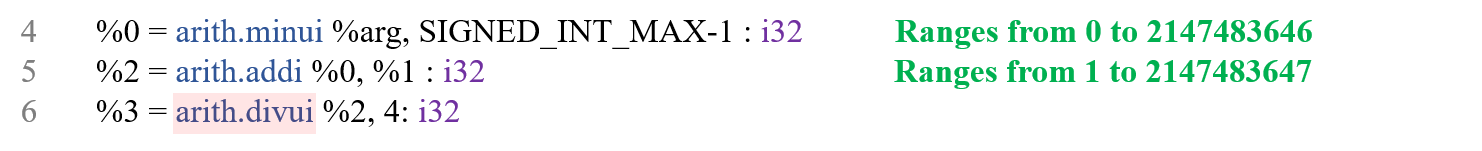}
  }

  \subfigure[MLIR program after utilizing undefined behavior elimination (for lines 2-3 in Figure~\ref{fig:moti1})]{
    \label{fig:moti4}
    \includegraphics[width=0.9\textwidth]{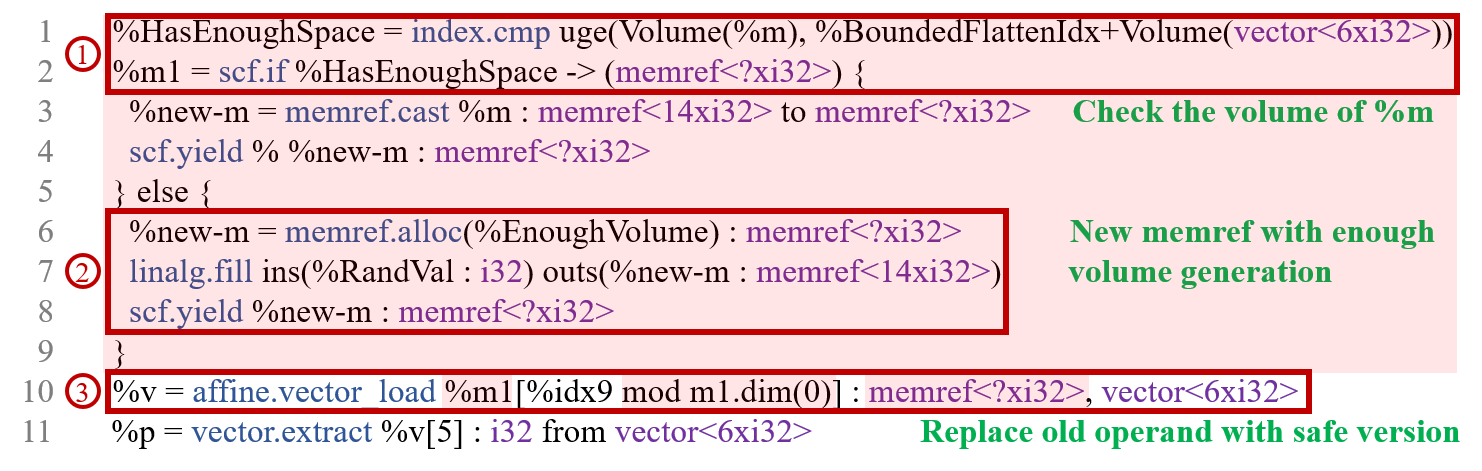}
  }

  \subfigure[Lowered MLIR program (for lines 10-11 in Figure~\ref{fig:moti2}) by applying ``-lower-affine'']{
    \label{fig:moti5}
    \includegraphics[width=0.9\textwidth]{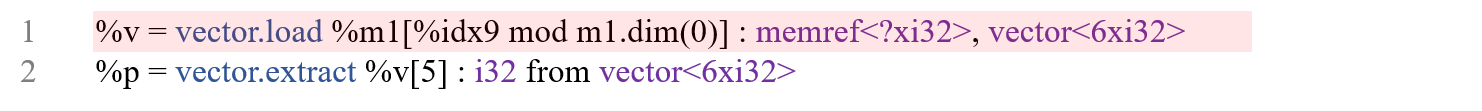}
  }

  \subfigure[Lowered MLIR program (for lines 10-11 in Figure~\ref{fig:moti2}) by applying ``-convert-vector-to-llvm'']{
    \label{fig:moti6}
    \includegraphics[width=0.9\textwidth]{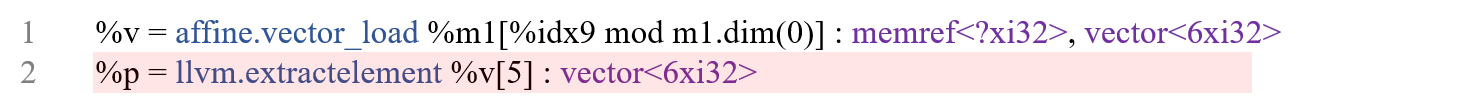}
  }
    \vspace{-3mm}
  \caption{Motivating example.}
  \label{fig:moti}
\end{figure}

\subsection{Terminology 
}


MLIR (Multi-Level Intermediate Representation) is a versatile and extensible intermediate representation designed to support multiple levels of abstraction and facilitate the development of various domain-specific compilers. 
To enable efficient compilation and optimization across different hardware and software targets,
MLIR introduces the concept of \textbf{dialects}, which are modular and extensible units that define custom operations, types, and attributes for specific domains or abstraction levels. 
An \textbf{operation} in MLIR is a fundamental unit of computation or transformation, representing a specific task or behavior within a dialect.
It takes as input a list of operands and attributes, performs a defined action, and produces one or more results as output.

For example, Figure~\ref{fig:moti1} illustrates an example of an MLIR program.
The program performs the following operations:
(1) Defines a constant integer value {\tt \%1} via {\tt arith.constant} operation (Line 1).
(2) Reads a vector {\tt \%v} from a memref value {\tt \%m} with a beginning position {\tt \%idx9} via {\tt affine.vector\_load} operation (Line 2).
(3) Extracts an element from the vector via {\tt vector.extract} operation (Line 3).
(4) Performs arithmetic calculations via \textit{arith} operations (Lines 4-6).
Specifically, the {\tt arith.addi} operation takes as input two {\tt i32} type operands ({\tt \%0} and {\tt \%1}), and produces a result of an {\tt i32} type value ({\tt \%2}) in line 5 in Figure~\ref{fig:moti1}.
Particularly, the attributes of an operation in MLIR actively participate in the computation process.
For example, the attribute {\tt value=1:i32} in the {\tt arith.constant} operation defines the literal value of the constant {\tt \%0} (Line 1 in Figure~\ref{fig:moti1}).

An MLIR program typically comprises operations from multiple dialects.
To compile it into an executable representation, these operations must be transformed into those within \textbf{target-specific dialects} (e.g., the \textit{llvm} and \textit{spirv} dialects, referred to as \textbf{executable dialects} for clarity in this paper).
This transformation enables program execution, which is essential for detecting silent bugs.
To achieve the transformation, MLIR provides a collection of lowering passes. 
Specifically, a \textbf{lowering pass} is a transformation that converts operations from one dialect to another, typically moving from higher-level abstractions to lower-level representations.
For example, the ``-convert-arith-to-llvm'' pass converts an \textit{arith} operations into an \textit{llvm} operation shown in Figure~\ref{fig:moti2}.
Additionally, MLIR provides a diverse set of \textbf{optimization passes}, each designed to enhance an MLIR program by improving performance, reducing resource consumption, or simplifying its structure while preserving its semantics.
These passes operate at various levels of abstraction and can be applied before or after lowering passes to refine the program and optimize execution efficiency.
For example, 
the ``-arith-unsigned-when-equivalent'' pass optimizes the program by replacing signed \textit{arith} operations to equivalent unsigned \textit{arith} operations.
For instance, 
in Figure~\ref{fig:moti1}, 
the signed division operation {\tt arith.divsi} in line 6 is replaced by the unsigned division operation {\tt arith.divui}, 
as both operands are positive, as shown in Figure~\ref{fig:moti3}.
Note that in our work, specifying lowering passes for a given MLIR program aims to compile it into an executable representation, ensuring successful execution.
In contrast, specifying optimization passes facilitates \textbf{cross-optimization differential testing} by generating multiple executable versions of the same MLIR program under different optimization strategies, helping to expose silent bugs.

Compiler bugs are generally categorized into two main types: crash bugs and silent bugs (also known as wrong code bugs)~\cite{DBLP:conf/pldi/YangCER11}.
Currently, no existing testing techniques are capable of detecting silent bugs in the MLIR compiler infrastructure, primarily due to the challenge posed by undefined behavior.
\textbf{Undefined behavior (UB)} refers to program constructs that result in unpredictable execution outcomes due to violations of a language's semantics or underlying constraints~\cite{DBLP:conf/pldi/YangCER11}. 
Unlike traditional programming languages, MLIR is an extensible compiler infrastructure with diverse dialects, each enforcing specific rules on operations, memory management, and data flow. 
UB can arise from various sources, such as uninitialized or out-of-bounds memory accesses, invalid type conversions, or violations of dialect-specific constraints (e.g., shape mismatches in \textit{tensor} operations).
For example, the {\tt affine.vector\_load} operation in line 2 of Figure~\ref{fig:moti1} demonstrates undefined behavior.
In this case, the remaining space in {\tt \%m} starting from index {\tt \%idx9} (value == 9) is 5,
which is insufficient to accommodate the required vector size of 6.
These issues present a major obstacle to silent bug detection in MLIR, as they can lead to non-deterministic behavior, masking actual compiler bugs or causing false positives during differential testing. 
Therefore, addressing UB is crucial to ensuring the reliability of MLIR-based compilation workflows and enabling effective silent bug detection.

\subsection{A Motivating Example}

Figure~\ref{fig:moti1} illustrates an MLIR program with undefined behavior.
Specifically, In Line 2,
the operation {\tt affine.vector\_load} attempts to read a value of type {\tt vector<6xi32>} from a memory reference value ({\tt memref<14xi32>}) {\tt \%m}, 
starting at position {\tt \%idx9} (value == 9).
Undefined behaviors may occur when the {\tt affine.vector\_load} operation encounters either of the following two conditions:
(1) 
Invalid index for the {\tt \%idx9}
: If {\tt \%idx9} exceeds the bounds of {\tt \%m}, undefined behavior will occur.
(2)
Insufficient remaining space for the {\tt \%m}:
If the remaining space in {\tt \%m}, starting from {\tt \%idx9}, is insufficient to accommodate the vector being loaded ({\tt \%v}), undefined behavior will also occur.
In this case, 
the operation requires space for 6 elements, but only 5 elements remain from {\tt \%idx9} in {\tt \%m}. This out-of-bounds access leads to undefined behavior, causing the loaded vector {\tt \%v} to contain unreliable values.
These unreliable values propagate through subsequent operations (e.g., {\tt vector.extract} in Line 3), 
ultimately affecting the execution result (assuming {\tt \%p} is printed). 
Such undefined behavior make all existing testing techniques hard to detect silent bugs due to unreliable execution results stemming from genuine optimization errors or undefined behaviors.
To detect silent bugs, it is essential to remove all undefined behavior in the MLIR programs.
Hence, \tech{} is proposed.
It solved this question by eliminating all undefined behavior and the updated program is shown in Figure~\ref{fig:moti4}, where modified code sections are highlighted in red. 
Specifically, \tech{} begins by 
inserting runtime checkers (marked by \ding{172} in Figure~\ref{fig:moti4}) that verify the volume of problematic memref {\tt \%m}.
These checks calculate both the memref's volume {\tt Volume(\%m)} and required vector space {\tt Volume(vector<6xi32>)}, then compare them with the flattened index after index-bounding ({\tt \%BoundedFlattenIdx}) to validate sufficient capacity. 
When insufficient space is detected, the system generates a safe operand 
(marked by \ding{173} in Figure~\ref{fig:moti4})
by allocating and initializing a properly-sized 
{\tt \%new-m};
otherwise, it preserves the original  {\tt \%m}
(Lines 3–4).
Finally (marked \ding{174}), all unsafe operands are replaced with their verified versions. \tech{} achieves this by replacing problematic operands {\%m} with {\%m1}, and applying index bounding through modulo arithmetic ({\tt \%idx9 mod m1.dim(0)}) to ensure memory safety while maintaining program semantics.
Through the above steps, \tech{} eliminates the undefined behavior, producing a UB-free MLIR program. This updated program is now suitable for differential testing.

After obtaining the UB-free program, 
another challenge is to lower the MLIR program into an executable form by
converting all dialects in the program into their executable forms. 
Since numerous operations may coexist within an MLIR program, 
and new dialects or operations can be generated during the lowering process, 
it is crucial for \tech{} to select an appropriate lowering path tailored to the given MLIR program.
For instance, 
consider the operations {\tt affine.vector\_load} and {\tt vector.extract} in Figure~\ref{fig:moti1}:
while {\tt vector.extract} can be directly lowered to \textit{llvm} using ``-convert-vector-to-llvm''(shwon as Figure~\ref{fig:moti5}), {\tt affine.vector\_load} first requires conversion via ``-lower-affine'' followed by ``-convert-vector-to-llvm''. That is, applying passes in the wrong order creates inefficiencies. Specifically, prematurely using ``-convert-vector-to-llvm'' leaves {\tt affine.vector\_load} unresolved (shwon as Figure~\ref{fig:moti6}), forcing redundant pass reapplications. The optimal approach first converts {\tt affine.vector\_load} to {\tt vector.load} form using ``-lower-affine'', then handles all \textit{vector} (i.e., {\tt vector.load} and {\tt vector.extract}) operations in a single ``-convert-vector-to-llvm'' pass. 
To find a suitable lowering path for given MLIR program, \tech{} further optimizes the lowering process by introducing a lowering-path optimization algorithm to convert all dialect in the given MLIR program into an executable form, which finally supports the differential testing.

\section{Approach}

In this section, we introduce the methodology of our approach, named \tech{}. It is the first technique that is specially designed for detecting silent bugs in the MLIR compiler infrastructure as far as we are aware. As introduced in Section~\ref{sec:intro}, \tech{} incorporates two major innovative components, i.e., \textbf{Undefined Behavior Elimination} (Section~\ref{sec:app:fixing}) and \textbf{Lowering Path Optimization} (Section~\ref{sec:app:pass}), to address the challenges of undefined behaviors and inefficiency of dialect lowering for generating executable MLIR programs. 

\begin{figure}[t]
    \centering
  \includegraphics[width=0.8\textwidth]
  {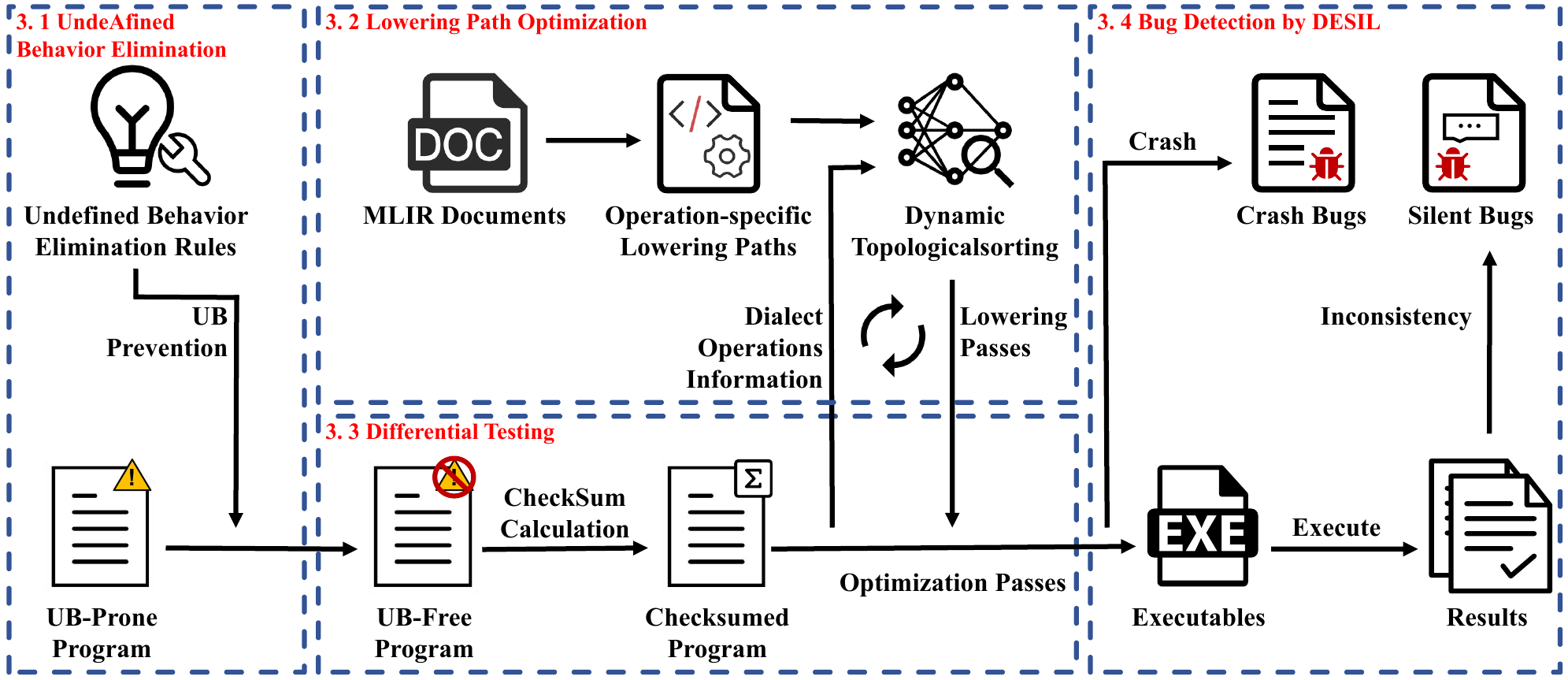}
  \vspace{-3mm}
    \caption{Overview of \tech{}}
    \label{fig:overview}
\end{figure}

Figure~\ref{fig:overview} presents the overall workflow of our approach. Specifically, \tech{} comprises a set of undefined behavior elimination rules, which effectively eliminate undefined behaviors for diverse UB-prone MLIR operations under certain conditions. Subsequently, to achieve lowering path optimization, we have defined a set of \textit{operation-specific lowering paths} in \tech{} for effectively transforming each dialect operation into the executable form through analyzing the corresponding documentation. By following this, \tech{} performs the lowering process via dynamically performing \textit{topological sorting} over all involved passes required by the dialect operations in the current MLIR program, thereby exploring the optimal \textit{lowering path} for efficient transformation.
Particularly, to evaluate whether the MLIR program is correctly compiled by the MLIR compiler, \tech{} leverages \textbf{differential testing} mechanism to detect inconsistent checksum values of the same MLIR program across different optimization sequences. We will introduce the detailed process in Section~\ref{sec:app:codeMod}.
Finally, we will outline the complete bug detection process in Section~\ref{sec:app:fuzz}.

\newcolumntype{x}[1]{>{\let\newline\\\arraybackslash\hspace{0pt}}p{#1}}
\renewcommand{\arraystretch}{1.2}

\begin{table}[]
\begin{threeparttable}
\caption{Undefined behavior elimination rules for different operations under certain conditions.}
\label{tab:strategy}
\scriptsize 
\begin{tabular}{x{2cm}|x{2cm}x{4.5cm}x{3.5cm}}
\toprule
\textbf{Operation Type} & \textbf{Example} & \textbf{Conditions}  & \textbf{UB Elimination Rules} \\ 
\hline
\midrule

\multicolumn{4}{c}{\footnotesize \textbf{UBs from Shape Inconsistency \tnote{$\dagger$}}} \\
\hline
\codeIn{affine.yield}.& 
\codeIn{affine.for iter\_args(\%arg0=\%m) affine.yield \%m1} & 
The operand (\codeIn{\%m1}) shape differs from argument (\codeIn{\%m}) shape of parent operation (\codeIn{affine.for}).  & 
Replace \codeIn{\%m1} with a new value has same shape as {\codeIn{\%m}}. \\
\hline

\codeIn{linalg.broadcast}.& 
\codeIn{linalg.broadcast ins(\%0) outs(\%1) dimensions=[1]} & 
The shapes of the two operands (\codeIn{\%0} and \codeIn{\%1}) are not same, except for the dimensions in \codeIn{dimensions}.  & 
Replace \codeIn{\%0} with a new tensor that has the same shape as \codeIn{\%1}, except for the  dimensions in \codeIn{dimensions}. \\
\hline

\codeIn{linalg.generic}.& 
\codeIn{linalg.generic \{ iterator\_types = ["p", "r"] \} ins(\%1, \%2, \%3) outs(\%4)} & 
\textbf{C1:} The shapes of \codeIn{ins} operands (\codeIn{\%1}, \codeIn{\%2}, \codeIn{\%3}) differ. \textbf{C2:} The dimension sizes of \codeIn{ins} operands (\codeIn{\%1}, \codeIn{\%2}, \codeIn{\%3}) specified as \codeIn{p} in \codeIn{iterator\_types} differ from the \codeIn{outs} operand (\codeIn{\%4}).  & 
For \textbf{C1}, ensure shapes of \codeIn{\%1}, \codeIn{\%2}, and \codeIn{\%3} are same. For \textbf{C2}, replace \codeIn{\%1}, \codeIn{\%2}, and \codeIn{\%3} with new tensors that have same shape as \codeIn{\%4} except for dimensions marked as \codeIn{r}. \\
\hline

\codeIn{linalg.matmul}.& 
\codeIn{linalg.matmul ins(\%1, \%2) outs(\%3)} & 
\textbf{C1:} The 2nd dimension of 1st \codeIn{ins} operand (\codeIn{\%1}) differs from the 1st dimension of 2nd \codeIn{ins} operand (\codeIn{\%2}). \textbf{C2:} The 1st dimension of 1st \codeIn{ins} operand (\codeIn{\%1}) differs from the 1st dimension of \codeIn{outs} operand (\codeIn{\%3}). \textbf{C3:} The 2nd dimension of 2nd \codeIn{ins} operand (\codeIn{\%2}) differs from the 2nd dimension of \codeIn{outs} operand (\codeIn{\%3}).  & 
Replace \codeIn{\%1} and \codeIn{\%2} with new tensors. For \textbf{C1}, ensure the 2nd dimension of \codeIn{\%1} equals to 2nd dimension of \codeIn{\%2}. For \textbf{C2}, ensure the 1st dimension of \codeIn{\%1} equals to 1st dimension of \codeIn{\%3}. For \textbf{C3}, ensure the 2nd dimension of \codeIn{\%2} equals to 2nd dimension of \codeIn{\%3}. \\
\hline

\codeIn{linalg.transpose}.& 
\codeIn{linalg.transpose ins (\%1) outs (\%2) permutation=[1, 0]} & 
The dimension of the \codeIn{ins} operand (\codeIn{\%1}) differs from the corresponding dimension of the \codeIn{outs} operand (\codeIn{\%2}) as specified in \codeIn{permutation}.  & 
Replace \codeIn{\%1} with a new tensor has same dimension as \codeIn{\%2} specified in \codeIn{permutation}. \\
\hline

Cast Operations (2).& 
\codeIn{\%m1 = memref.cast \%m0 : memref<?xi32> to memref<10xi32>} & 
The static dimension of result (\codeIn{\%m1}) differs from corresponding runtime dynamic dimension size of operand (\codeIn{\%m0}).  & 
Replace \codeIn{\%m0} with new memref with dynamic dimension equals to corresponding static dimension of \codeIn{\%m1}. \\
\hline

Operations with same shape operands (3).& 
\codeIn{linalg.copy ins(\%0) outs(\%1)} & 
The shapes of two operands (\codeIn{\%0} and \codeIn{\%1}) differ.  & 
Replace \codeIn{\%0} with new tensor that has the same shape as \codeIn{\%1}. \\
\hline



\midrule

\multicolumn{4}{c}{\footnotesize \textbf{UBs from Index Out-of-Bounds}} 
\\
\hline
Scalar Value Load and Store Operation (8) & 
\codeIn{affine.store \%0, \%m[\%idx1, \%idx2]} & 
Any index (\codeIn{\%idx1}, \codeIn{\%idx2}) exceeds the dimensions of the array-like operand (\codeIn{\%m}). & 
Confine the index values to the dimensions of \codeIn{\%m} using \codeIn{index.remu}. \\

\hline

Dim Operations (2) & 
\codeIn{tensor.dim \%t, \%0} & 
The index (\codeIn{\%0}) exceeds the rank of array-like value (\codeIn{\%t}). & 
Confine \codeIn{\%0} within the rank of \codeIn{\%t}. \\

\hline

Array-Like Value Store and Load Operations (2) \tnote{$\dagger$} &\codeIn{affine.vector\_store \%o,\%c[\%idx0,\%idx1]}  & 
\textbf{C1}: Any index (\codeIn{\%idx0}, \codeIn{\%idx1}) exceeds the dimensions of the data container (\codeIn{\%c}). \textbf{C2}: The remaining space in container (\codeIn{\%c}) starting from position (\codeIn{\%idx0}, \codeIn{\%idx1}) is insufficient to load/store the object (\codeIn{\%o}). & 
For \textbf{C1}, Confine the index to the dimensions of \codeIn{\%c} using \codeIn{index.remu} operation. For \textbf{C2}, replace the container (\codeIn{\%c}) with a new one with sufficient space.  \\
\hline
\midrule

\multicolumn{4}{c}{\footnotesize \textbf{UBs from Invalid Memory References}} \\
\hline
\codeIn{memref. assume\_alignment}.\tnote{$\dagger$} & 
\codeIn{memref.assume\_ alignment \%m, 4} & 
The alignment attribute (\codeIn{4}) differs from the origianl alignment information of the operand (\codeIn{\%m}).  & 
Replace alignment attribute with alignment information of \codeIn{\%m}, or default value if \codeIn{\%m} has no information. \\
\hline

\codeIn{memref.realloc}. & 
\codeIn{\%m1 = memref. realloc \%m0} & 
The original memref value (\codeIn{\%m0}) is used after this operation.  & 
Replace the use of \codeIn{\%m0} after this operation with a new memref value. \\
\hline

\codeIn{Memory Allocation Operations}. & 
\codeIn{\%0 = memref.alloca () : memref<1xi32>} & 
Directly use the value in the result memref value without initialization.  & 
Initilize the content of memref value with linalg.fill. \\
\hline

\midrule

\multicolumn{4}{c}{\footnotesize \textbf{UBs from Scalar Calculations}} \\
\hline
Shift Operations (6).& 
\codeIn{index.shrui \%1, \%2} & 
The 2nd operand (\codeIn{\%2}) exceeds the bit width of the 1st (\codeIn{\%1}).  & 
Replace \codeIn{\%2} with a random constant value within the bit width of \codeIn{\%1}. \\

\hline

Signed Integer Division Operations (6)& 
\codeIn{index.divs \%1, \%2} & 
\textbf{C1}: The 2nd operand (\codeIn{\%2}) is zero. \textbf{C2}: The 1st operand (\codeIn{\%1}) is \codeIn{INT\_MIN} (specific to its bitwidth), and the 2nd operand (\codeIn{\%2}) is -1.  & 
For \textbf{C1}, make \codeIn{\%2} unequal to 0. For \textbf{C2}, make \codeIn{\%2} unequal to -1.  \\

\hline

Unsigned Integer Division (4) and Remainder Operations (4) & 
\codeIn{index.divu \%1, \%2} & 
The 2nd operand (\codeIn{\%2}) is zero. & 
Replace \codeIn{\%2} with nonzero random signed or unsigned integer value (according to type of \codeIn{\%2}).  \\


\bottomrule
\end{tabular}

\begin{tablenotes}
\footnotesize
\item[$\dagger$] {\scriptsize The conditions and elimination rules of these undefined behaviors are MLIR-specific.}
\end{tablenotes}

\end{threeparttable}
\end{table}

\subsection{Undefined Behavior Elimination}
\label{sec:app:fixing}


Undefined behaviors (UBs) in MLIR programs for testing the MLIR compiler infrastructure can lead to unpredictable execution results, making it difficult to accurately detect silent bugs since it is hard to determine whether the unexpected execution results of the MLIR programs are induced by their inherent UBs or the silent bugs in the compiler. Therefore, eliminating the undefined behaviors is essential. However, ensuring that the generated MLIR programs are UB-free is challenging due to the diverse root causes of such behaviors, especially those that are specific to MLIR programs. For instance, MLIR programs may easily cause UBs that are due to the shape (or dimensions) inconsistency while involving array-like values (e.g., tensor), requiring effective UB elimination methods tailored to handle such cases.
To address this challenge, we conducted a comprehensive analysis of those operations that are supported and frequently used by existing MLIR fuzz techniques and summarized the potential UBs they may induce by carefully examining their usage documentation. 
Specifically, we refer to operations that may cause undefined behaviors as \textit{UB-prone} operations. Based on the conditions that trigger potential UBs for each UB-prone operation, we manually defined a set of undefined behavior elimination rules to modify programs and ensure that UBs cannot be triggered. The details of the undefined behavior elimination rules are presented in Table~\ref{tab:strategy}. In this table, we list the types of UB-prone operations, followed by an example to clearly present the conditions for triggering potential UBs, and then we summarized the undefined behavior elimination rules to eliminate the trigger of the UBs. In particular, one operation type may involve multiple UB-prone operations, which share similar root causes and undefined behavior elimination rules. The number in the brackets shown in the first column indicates the number of involved operations belonging to the specific operation type. For clarity, we only present one representative example in the table to aid the understanding and illustration. The complete operations and their associated undefined behavior elimination rules can be found at our project's homepage~\cite{DESILrepository}.

Consequently, given an MLIR program for testing MLIR compilers, \tech{} first identifies all UB-prone operations within it. For each identified operation, \tech{} applies the corresponding undefined behavior elimination rule to generate the correct program. It is important to note that since the lowering process (as discussed in Section~\ref{sec:app:pass}) should maintain the semantics of the MLIR program, i.e., UB-free programs should not encounter any UBs after the transformation.
As a consequence, the fix process is a one-off task for each MLIR program. By sufficiently fixing all potential UBs in the initial MLIR programs, our approach ensures comprehensive mitigation and ensures that UBs cannot be triggered in the target executable programs. In the following, we will provide a proof-of-concept introduction of these undefined behavior elimination rules. The detailed implementations for fixing each UB can be found in our open-source repository.


\subsubsection{Undefined Behaviors from Shape Inconsistency}



This type of UBs is typically due to the calculation related to vector-like values, such as matrices or tensors that are usually involved in deep learning programs. These UBs are usually triggered because the shapes of two tensors (or dimensions of matrices) do not match each other. Actually, this type of UBs is typically specific to MLIR programs due to their frequent use of array-like values.
In contrast, traditional programming languages typically decompose such
operations into loops and scalar value computations, and thus are free from this type of UBs.
For example, two tensor values {\tt [1,2]} and {\tt [1,2,3]} cannot perform multiplication since their shapes (or dimensions) are unmatched. Specifically, we summarized three situations where the shape inconsistency may cause potential UBs: (1) shape inconsistency between arguments and return values, such as the operation of {\tt linalg.matmul ins(\%1,\%2) outs(\%3)} requiring the dimensions of arguments (i.e., {\tt \%1} and {\tt \%2}) and the return value (i.e., \%3) to match each other;
(2) Shape inconsistency in a specified dimension, such as the argument dimension in {\tt linalg.transpose} is inconsistent with the output dimension specified by {\tt permutation=[1,0]}; (3) Shape inconsistency between source and target operand, such as {\tt linalg.copy} should not change the value dimension during copying.

To address these inconsistencies, our approach will insert shape-related runtime checking code as the checker of memory allocation explained above, and generate suitable operands for replacement if any inconsistency was found.
It is important to note that, in general, \tech{} avoids modifying the shape of the return value of operations, as it tends to affect all follow-up uses of the result value, and thus increases the risk of the modified MLIR program being rejected by the MLIR front-end due to checks related to shape. As a consequence, \tech{} will always update the shapes of the others except for the return value. Different from the above situations, the {\tt tensor.empty} operation is commonly used by some MLIR fuzz testing techniques (e.g., MLIRSmith) for generating MLIR programs. However, as explained in the corresponding documentation, this operation may cause unpredictable results since its values are unpredictable. To avoid UBs induced by it, \tech{} replaces all appearance of {\tt tensor.empty} with either {\tt tensor.from\_elements} or {\tt tensor.splat} for initializing new tensors.

\subsubsection{Undefined Behaviors from Index Out-of-Bounds}



This type of undefined behavior is prevalent across various programming languages, 
occurring when an MLIR program attempts to access a memory location exceeding the bound of a valid range. This UB is critical as it always results in unreliable execution results and execution crashes, and thus should be eliminated. Like many other programming languages, this kind of UBs in MLIR programs usually happen in two scenarios:
(1) accessing an array-like value with the specified index (e.g., {\tt tensor.dim}), (2) and storing an object to a data container without sufficient available memory space (e.g., {\tt affine.store}).
For the first scenario, 
our straightforward idea for fixing is to check whether the given {\tt index} exceeds the range of the array-like value, and then replace the index with a valid value within the range. For the second scenario, 
the undefined behavior elimination rule is to allocate another memory to ensure the available space is sufficient for storing the object. It is important to note that this scenario is specific to MLIR programs due to its high-level abstraction of data types~\cite{affine}.  

In particular, checking whether the remaining memory is sufficient is not statically doable since the memory will be dynamically allocated and consumed during the running of the MLIR program. Therefore, to ensure the fix is valid and effective, sometimes we are expected to insert new code logic for dynamically checking the triggering conditions of certain UBs and eliminate them on demand. For example, as shown in Table~\ref{tab:strategy}, the operation ({\tt affine.vector\_store \%o,\%c[\%idx0,\%idx1]}) is designed to store a vector object (i.e., {\tt \%o}) into the data container {\tt \%c} starting from position {\tt [\%idx0,\%idx1]}. In this case, to ensure the store operation is correctly performed, our approach will insert multiple lines of code for dynamically checking the size of {\tt \%o} and the available memory of {\tt \%c}, and allocate additional memory if needed. In this way, 
memory is guaranteed to meet the requirement during the execution of the MLIR program, and thus the undefined behavior can be avoided.

\subsubsection{Undefined Behaviors from Invalid Memory References.}

This type of UBs is primarily caused by the invalid references to memory, and the associated operations are commonly from the \textit{memref} dialect. Similarly, these UBs may cause crashes during running the MLIR program or produce unpredictable results. In summary, the root causes of these UBs are twofold: (1) conflict between actual memory alignment and specified alignment assertions by using the operation {\tt memref.assume\_alignment}; (2) reference to invalid memory, such as accessing uninitialized or reallocated memory by {\tt memref.realloc}.
Regarding the first root cause,
\tech{} will update the specified alignment attribute in the assertion operation and make it align with the actual value. Regarding the second root cause, \tech{} incorporates a define-use chain analysis~\cite{DBLP:books/aw/AhoSU86, DBLP:journals/toplas/HarroldS94} for identifying the invalid memory access, and then updates the invalid references to a valid one or initializes the referenced memory directly. In particular, to avoid out-of-memory crashes caused by the memory allocation operations (e.g., {\tt memref.alloc} and {\tt memref.alloca}) during continuously allocating memory space, \tech{} restrains the dimensions of an array-like operand not exceeding 4$\times$32.


\subsubsection{Undefined Behaviors from Scalar Calculations.}







This type of undefined behaviors are primarily caused by the operations related to scalar value calculations. In particular, it mainly includes three root causes -- shift overflow (e.g., {\tt index.shrui}), signed integer overflow (e.g., {\tt index.divs}), and division by zero (e.g., {\tt arith.ceildivsi}). Effectively eliminating this kind of UBs is crucial since they always cause crashes or unpredictable execution results while executing the compiled MLIR program, disabling the precise detection of silent bugs. To address these UBs, we have defined a viable undefined behavior elimination rule for each kind of root cause (as presented in Table~\ref{tab:strategy}). For example, in the division operations (e.g., {\tt arith.ceildivsi}), undefined behavior will arise due to division by zero or signed division overflow (e.g., dividing the minimum signed integer value by -1). To fix this, we first check whether the divisor operand in this operation is zero or not through either static or dynamic analysis, and then replace zeros with a randomly generated integer value unequal to zero.
Similarly, in shift operations (e.g., {\tt index.shl} and {\tt arith.shli}), 
if the second operand (i.e., the bits of shifting) exceeds the bitwidth of the initial value (i.e., the first operand), 
an unpredictable value will be returned. In this case, the undefined behavior elimination rule is to confine the value of the second operand within the bitwidth of the first operand. In this way, the UBs can be effectively avoided. Different from existing methods (e.g., CSmith), typically adopting predefined safe wrapper functions, for avoiding UBs from scalar calculations, \tech{} directly seeds the UB checking and elimination logic into the initial MLIR program. As a consequence, our approach can effectively reduce the code size compared to existing methods by solely generating relevant elimination logic for used bitwidths, which significantly improves the lowering efficiency (will be introduced in Section~\ref{sec:app:pass}) by avoiding much irrelevant code involved by predefined wrapper functions.

\subsubsection{UB-Irrelevant Fix for Normal Compilation.}
\label{sec:ubirefix}
Besides preventing undefined behaviors presented above, 
there is another unique case -- 0-dimension objects -- that arises from MLIR's rich semantics supporting 0-dimensional constructs some array-like data types such as tensors and memrefs.
This may cause the dialect lowering process (i.e., compilation) failed since some low-level dialects (e.g., {\tt vector}) do not support the dimension of objects to be zero. As a consequence, \tech{} further incorporates an additional undefined behavior elimination rule for such cases. Specifically, \tech{} replaces 0-dimension objects with non-0-dimension objects to ensure the MLIR program can be successfully transformed into the executable ones.

\begin{algorithm}[t]
  \small
  \SetAlgoLined
  \SetKwInOut{Input}{Input}
  \SetKwInOut{Output}{Output}
  \Input{
      \textit{Program}: an MLIR program that may contain UB-prone operations\\
  }
  \Output{
      \textit{Program}: an UB-free MLIR program after applying undefined behavior elimination rules\\
  }

  \SetKwFunction{FFixUB}{FixUB}
  \SetKwProg{Cn}{Function}{:}{\KwRet \textit{Program}}
  \Cn{\FFixUB{$Program$}}{    
  {\scriptsize\tcc{Collect all UB-prone operations from the top level operation of \textit{Program}.}}
  \textit{UBProneOps} = \textit{CollectUBProneOps}(\textit{Program}.\textit{TopLevelOp})  \\
    \ForEach {op in UBProneOps} {
      {\scriptsize\tcc{Eliminate potential UBs in each operation by applying the associated undefined behavior elimination rule.}}
      $program=UBEleminationForOp(program, op)$   \\
  }
  {\scriptsize\tcc{Conduct UB-Irrelevant fix for normal compilation.}}
  $program = UBIrrelevantFix(program)$
  
  }

  \vspace{1mm}
  \SetKwFunction{FUBEleminationForOp}{UBEleminationForOp}
  \SetKwProg{Wn}{Function}{:}{\KwRet \textit{Program}}
  \Wn{\FUBEleminationForOp{$program$, $op$}}{    
    {\scriptsize\tcc{Insert runtime checkers to the MLIR program for detecting and avoiding potential UBs.}} 
    $program = InsertRuntimeCheck(program, op)$ \\ 
    {\scriptsize\tcc{Generate the safe version of operands that eliminate the UB based on the runtime checker.}} 
    $program, safeOp = PrepareSafeOperands(program, op)$ \\
    {\scriptsize\tcc{Replace the old UB-prone operands with corresponding UB-free operands.}}
    $program = ReplaceOldOperandWithUBFree(program, safeOp)$ 
  }

  \vspace{1mm}
  \SetKwFunction{FCollectUBProneOps}{CollectUBProneOps}
  \SetKwProg{Pn}{Function}{:}{\KwRet \textit{UBProneOps}}
  \Pn{\FCollectUBProneOps{$operation$}}{
    \textit{UBProneOps} = [];  \\
   
   {\scriptsize\tcc{Recursively collect the UB-prone operations if the \textit{operation} has Block.}}
   \If{operation.hasBlock()} {
    \For{o in operation.getBody()} {
      \textit{UBProneOps} = \textit{UBProneOps} $\bigcup$ \textit{CollectUBProneOps}(\textit{operation})
    }
   } 

   {\scriptsize\tcc{Collect UB-prone operations that may cause UBs according to its operation type.}}
   \If {IsUBProneOpType(operation)} {
    \textit{UBProneOps}.\textit{add}(\textit{operation})
   }
  }
  
  \caption{Undefined Behavior Elimination}
  \label{alg:ubfix}
\end{algorithm}

\textbf{UB Elimination Algorithm}: Based on the undefined behavior
elimination rules introduced above, given an MLIR test program, \tech{} tries to fix all potential UBs in it by following the process presented in Algorithm~\ref{alg:ubfix}.
In general, the algorithm takes an MLIR program that may contain UB-prone operations as the input, and outputs a new MLIR program that are expected to be UB-free by applying the necessary undefined behavior elimination rules explained above. Specifically, given the input MLIR \textit{Program}, \tech{} first collects all UB-prone operations in it (Line 2). In particular, 
an MLIR program compromises a set of operations, which are typically structured as recursively nested code regions like traditional programming languages (e.g., {\tt While} statements usually include other statements in a Java or C++ program). That is, an operation may have one or multiple nested regions (known as Blocks)~\cite{MLIR_language_reference}, each of which will also consists of a list of operations. By following this structure, \tech{} recursively traverses the MLIR program in a top-down fashion (\textit{Program.TopLevelOp} indicates the outmost operation in the program) for collecting all UB-prone operations within the program (Lines 15-22). Then, for each UB-prone operation (Line 3), \tech{} tries to eliminate the potential UBs by applying the associated undefined behavior elimination rules (Line 4) by calling the function of \textit{UBEliminationForOp()} (Lines 8-12). Specifically, \tech{} inserts necessary runtime checkers into the MLIR program for checking whether any conditions associated to the UB-prone operation are satisfied during the MLIR program execution (Line 9), and then generates safe operands based on the checking results and the associated undefined behavior elimination rules (Line 10) for replacing the original operands used in the operation (Line 11). For example, to fix the potential UBs from index-out-of-bounds errors caused by 
operation {\tt affine.vector\_store \%o,\%c[\%idx0,\%idx1]}, \tech{} inserts several lines of code (i.e., runtime checker) to the MLIR program for check whether the available memory in {\tt \%c} is sufficient to store the object {\tt \%o}. If it is insufficient, \tech{} inserts new operand (e.g., \%d) with sufficient memory in the MLIR program. Finally, \tech{} replaces the original operand {\tt \%c} with {\tt \%d}. The overall process concludes with the undefined behavior-irrelevant fixes for normal compilation, as introduced in Section~\ref{sec:ubirefix}. In this way, the modified MLIR program will include those runtime checkers that can effectively avoid the trigger of potential UBs during the execution of the MLIR program.

\subsection{Lowering Path Optimization}
\label{sec:app:pass}

Given an MLIR program, after eliminating the undefined behaviors in it presented above, the next process is to transform the program into executable form. As aforementioned, this transformation process is indeed challenging because MLIR programs usually contain operations from diverse dialects at different levels, and each dialect operation may require a specific sequence of lowering passes -- \textit{operation-specific lowering path} -- before reaching a fully executable form. Furthermore, new dialect operations will be dynamically produced during the lowering process, aggravating the difficulty of this process. However, exhaustively enumerating all possible sequences of lowering passes is impractical since it would impose a significant efficiency bottleneck, and thus affect the effectiveness of silent bug detection.

\begin{figure}[t]
    \centering
  \includegraphics[width=0.7\textwidth]{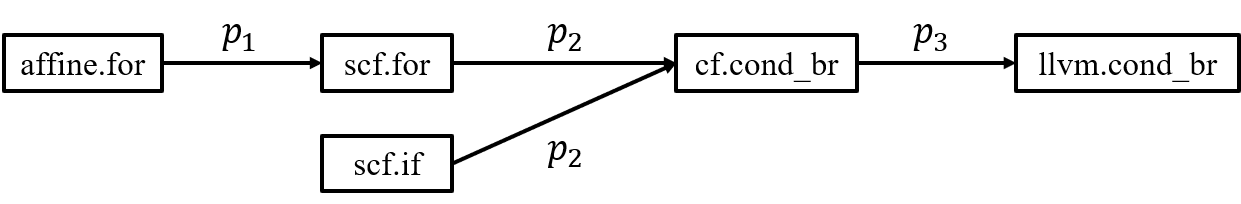}
    \caption{The lowering process for the dialect operation {\tt affine.for} by applying passes $\{p_1,p_2,p_3\}$.}
    \label{fig:conversion}
\end{figure}

To address this challenge, \tech{} incorporates an innovative lowering-path optimization algorithm that dynamically determines the optimal lowering pass based on the dialect operations included in the MLIR program. Specifically, \tech{} first builds a mapping between lowering passes and dialect operations based on the MLIR documentation, recording an \textit{operation-specific lowering path} for each operation. Then, given an MLIR program, \tech{} determines its optimal lowering path by performing \textit{topological sorting} on the lowering passes derived from the operation-specific lowering paths of the program’s
operations. In this way, \tech{} can efficiently transform a given MLIR program into the executable form by avoiding the circular application of the same lowering passes. 
In summary, the MLIR program compilation (or lowering path optimization) process in \tech{} consists of two key stages which are presented as follows:
\begin{enumerate}
    \item \textbf{Operation-Specific Lowering Path Construction}: To ensure that every operation in the MLIR program can be successfully transformed into the executable form, \tech{} builds the mapping between each dialect operation and a sequence of lowering passes,
    which can transform the associated operation into the executable form. 
    Specifically, we call such a sequence of lowering passes \textit{operation-specific lowering path}. Formally, it is defined as a tuple of $\left\langle o, P, R \right\rangle$, where $P=\{p_1, p_2, \cdots,p_n\}$ is a set of lowering passes that are needed to transform the operation $o$ into the executable form, and $R=\{p_i\succ p_j| p_i,p_j\in P\}$ defines the partial order between two lowering passes, specifying the pass $p_i$ in $P$ should be executed before $p_j$ for transforming $o$. Figure~\ref{fig:conversion} presents an example of the lowering process for the dialect operation {\tt affine.for}, which will be transformed into the executable {\tt llvm.cond\_br}. Consequently, the \textit{operation-specific lowering path} associated to the operation {\tt affine.for} is $\left\langle \mathrm{{\tt affine.for}}, \{p_1, p_2, p_3\}, \{p_1\succ p_2,p_2\succ p_3\}\right\rangle$. Specifically, the output of this stage is the \textit{operation-specific lowering path} for each operation.
    In particular, to ensure the reliability of mapping results, we manually analyzed the documentation of operations and lowering passes. Moreover, we verified all the operation-specific lowering paths by constructing associated MLIR programs to ensure they actually work.

    \item \textbf{Lowering Pass Topological Sorting}: Given that MLIR programs may include a variety of operations and each of them is associated with an \textit{operation-specific lowering path}. To achieve an efficient lowering process (i.e., compilation) and avoid circular application of the same lowering pass, \tech{} exploits the optimal pass execution by leveraging \textit{topological sorting} over all the passes associated to all the operations in the current MLIR program. 
    Formally, assuming $O$ represents all the operations in the program, and $P$ is all passes involved. Then, $\forall p_1, p_2\in P$, $p_1\succ p_2$ holds \textit{iff} it holds for any $o\in O$. According to this, \tech{} always chooses the pass $p_i\in P$ as the first one for execution \textit{iff} $\nexists p_j\in P, p_j\succ p_i$. It is important to note that such a pass $p_i$ always exists since each dialect operation is guaranteed to be transformed into the executable form by the corresponding \textit{operation-specific lowering path}, indicating no circular dependency exists for all the passes. However, if more than one pass meets the condition, \tech{} randomly chooses one of them. Figure~\ref{fig:conversion} presents such an example, where the lowering pass $p_1$ will be the first lowering pass for execution.
\end{enumerate}

In summary, the first stage (i.e., Operation-Specific Lowering Path Construction) is a one-off task. Once the operation-specific lowering path are constructed, they can be directly reused during the compilation process for diverse MLIR programs. In contrast, the second stage will be dynamically performed by \tech{} -- choosing a lowering pass to execute for transforming the MLIR program into another form -- until all the operations in the program are transformed into the executable form.


\subsection{Differential Testing}
\label{sec:app:codeMod}

After transforming the given MLIR program into an executable form, the next step is to run the program and examine whether it was correctly compiled by the MLIR compiler. However, like all fuzz testing techniques, it is infeasible to automatically obtain the oracles of the test execution without the specification of the test program~\cite{DBLP:conf/ecoop/Xie06}. 
To address this issue, 
\tech{} employs differential testing to check whether the MLIR program is correctly compiled by the compiler. To achieve that, \tech{} comprises a compilation operation-aware optimization recommendation strategies along with the lowering pass topological sorting (introduced in Section~\ref{sec:app:pass}). The objective of this strategy is to produce different test execution results by applying diverse optimizations during compiling the same MLIR program, where potential silent bugs in the MLIR compiler would be detected.
Specifcially, given an MLIR program, \tech{} collects all the operations involved in it and then selects \optnum{} (which is evaluated in Section~\ref{sec:param}) compiled programs according to the collected operations.
Then, \tech{} checks the execution results of these different executions (\tech{} by default generates \diffnum{} executable programs for each MLIR program). Different execution results among them indicate incorrect compilations, and represent the detection of silent bugs in compilers.

However, unlike traditional test programs for high-level programming languages (e.g., Java and C++), which usually associate with a relatively complete functionality, the MLIR programs may not perform a meaningful function as they are usually generated in a random fashion by assembling low-level dialect operations. As a consequence, the final output of the MLIR programs may not well reflect their whole execution behaviors. This issue potentially reduces the detection capability of silent bugs in MLIR compilers since the mis-compiled operations may not affect the final execution results. To improve the capability of the MLIR program for detecting silent bugs, \tech{} further incorporates an ``test oracle'' generation process inspired by Csmith~\cite{DBLP:conf/pldi/YangCER11} -- A well-known fuzz testing method for C/C++ compilers. Specifically, \tech{} calculates the checksum of the MLIR program by summing up all the accessible integer values (values in array-like objects are also included) at the end of the MLIR program's main function. In particular, \tech{} does not consider floating-point values since the precision issue during calculation may cause false positives in bug detection.
Finally, the checksum will play as an estimation of the test ``oracle''. Since any integer value error will propagate to the final checksum, it should have a strong power to uncover incorrect execution results, thereby improving the capability of detecting silent bugs in MLIR compilers.

\subsection{Bug Detection by \tech{}}
\label{sec:app:fuzz}

Given an MLIR program, \tech{} first transforms it to the executable forms by adopting different sequences of optimizations. Then, it runs the test programs and compares the checksum correspondingly, and reports potential silent bugs if the checksum values are inconsistent. More specifically, the overall silent bug detection process of \tech{} consists of four stages, which are presented as follows.

\begin{enumerate}
    \item \textbf{Test Program Generation}: In order to generate diverse MLIR programs for detecting silent bugs in MLIR compilers, \tech{} utilizes a MLIR program generator. It is important to note that \tech{} is a post processing method for MLIR program generators, and thus can be combined with any off-the-shelf generators as a plugin.
    
    \item \textbf{Undefined Behavior Elimination}: For each candidate MLIR program, \tech{} fixes the undefined behaviors in it by leveraging the UB elimination algorithm presented in Section~\ref{sec:app:fixing}. As aforementioned, this process is essential to ensure the capability of precisely detecting silent bugs since UBs are easy to produce false positives.

    \item \textbf{Lowering Path Optimization}: After eliminating potential UBs in the MLIR programs, \tech{} leverages the lowering path optimization component (introduced in Section~\ref{sec:app:pass}) to transform the programs into executable forms. 

    \item \textbf{Differential Testing for Bug Detection}: 
    Given different executable programs compiled from the same MLIR program, \tech{} generates the calculations of the checksum for each one. 
    For differential testing, \tech{} leverages the operation-aware optimization recommendation component for generating different versions of the MLIR programs. 
    Finally, \tech{} executes the programs and detects potential silent bugs by checking the consistency of their associated checksum values. In particular, \tech{} also has the ability to detect crash bugs in MLIR compilers during the compilation process if any crashes are encountered.
\end{enumerate}

\section{Evaluation}
To evaluate \tech{}, we designed the following research questions (RQs) in the study:
\begin{itemize}
\item RQ1: How effective is \tech{} in detecting previously unknown MLIR bugs?
\item RQ2: How does \tech{} perform compared to the state-of-the-art MLIR testing techniques?
\item RQ3: How does each component contribute to the overall effectiveness of \tech{}?
\item RQ4: What is the influence of different configurations on the effectiveness of \tech{}?
\end{itemize}

\subsection{Experimental Setup}

To answer RQ1, we applied \tech{} to fuzz the latest versions of the MLIR compiler infrastructure, aiming to uncover previously unknown bugs.
Over a four-month fuzzing period, we consistently updated the infrastructure to the latest version, covering 
revisions from {\tt adbf21} to {\tt b6d5fa}.
To answer RQs 2-4, we selected the latest version of the MLIR compiler infrastructure at the time of performing these experiments (i.e., revision {\tt c6d6da}).
We ran each studied technique for 12 hours on this version.
To reduce the influence of randomness involved in testing, we repeated each technique for five times (except the variant techniques investigated in RQ4) and reported the aggregated results.
Due to the large number of studied variant techniques in RQ4 and the fuzzing cost for each technique, we repeated them for three times to balance the conclusion robustness and evaluation cost, and then reported the aggregated results.

By default, we set the number of optimization passes per lowering step (\optnum{}) to 1 and the number of compilations for differential testing (\diffnum{}) to 2 in \tech{} for seeking cost-effectiveness.
The influence of different settings for them will be investigated in RQ4.
All our experiments were conducted on a machine with Intel(R) Xeon(R) CPU E5-2640 v4 @ 2.40GHz and 128G Memory, Ubuntu 20.04.6 LTS.

\subsubsection{Studied Techniques}\label{sec:baseline}
Due to the pluggable design of \tech{} presented before, \tech{} can be combined with any existing MLIR program generation tools.
That is, for any given MLIR program, \tech{} can be applied to transform it into a UB-free one and then compile it to the executable program for testing.
In the study, to ensure the generalizability of \tech{}, we used two state-of-the-art MLIR program generation tools (i.e., MLIRSmith~\cite{DBLP:conf/kbse/WangCXLWSZ23}
and MLIRod~\cite{DBLP:conf/issta/Suo00JZW24}
) to prepare initial MLIR programs for \tech{}, respectively.
For ease of presentation, we call the two instantiations \textbf{\tech{}$_{\textit{smith}}$} and \textbf{\tech{}$_{\textit{od}}$}.

Since the MLIR programs generated by MLIRSmith and MLIRod may contain undefined behaviors, neither includes a lowering component to transition these programs into executable forms. 
As a result, their original versions cannot detect silent bugs.
To enable a more comprehensive comparison, we integrated lowering path optimization and differential testing components from \tech{} into MLIRSmith and MLIRod, equipping them with the capability to detect silent MLIR bugs. 
To distinguish these enhanced versions from their originals, we refer to them as \textbf{MLIRSmith$_{\textit{enhanced}}$} and \textbf{MLIRod$_{\textit{enhanced}}$}, respectively.
Specifically, these variants retain their original program generation mechanisms but follow \tech{}’s compilation process to produce executable programs for differential testing.
However, since MLIRSmith$_{\textit{enhanced}}$ and MLIRod$_{\textit{enhanced}}$ do not eliminate UB, they may produce a high number of false positives in silent bug detection. 
Comparing them against \tech{} allows us to evaluate RQ2 and also demonstrates the contribution of the UB elimination component in \tech{}.



Besides the above-mentioned UB elimination component, there are another two main components in \tech{} - the lowering path optimization and differential testing components. 
Their contributions will be investigated in RQ3.
To investigate the contribution of the lowering path optimization component, we constructed the corresponding variants \textbf{\tech{}$_{\textit{smith}}^{\textit{w/o lower}}$} and \textbf{\tech{}$_{\textit{od}}^{\textit{w/o lower}}$} by removing this component from \tech{}.
Specifically, these variants randomly select a sequence of lowering passes to construct a lowering path for each MLIR program after UB elimination. 
To prevent the lowering process from hanging due to the application of an excessive number of passes, we set the lowering path length to 50, which is significantly larger than the average length of the determined lowering paths in \tech{} during our study.

Regarding the differential testing component, \tech{} modifies the application of optimizations to generate a set of executable programs for differential comparison.
To enhance the effectiveness of differential testing, it employs a recommendation mechanism that selects optimization passes based on the operations present in the MLIR program, rather than choosing them randomly. 
This mechanism increases the likelihood of optimizations affecting the program's behavior, improving the chances of exposing silent bugs.
Therefore, in RQ3, we also investigated the contribution of this optimization recommendation mechanism by constructing the corresponding variants \textbf{\tech{}$_{\textit{smith}}^{\textit{w/o opt}}$} and \textbf{\tech{}$_{\textit{od}}^{\textit{w/o opt}}$}, which remove this mechanism from \tech{}$_{\textit{smith}}$ and \tech{}$_{\textit{od}}$ respectively.
Specifically, these variants randomly select optimization passes to generate a set of executable programs for differential testing, rather than leveraging operation-aware recommendations.

To answer RQ4, we configured the number of optimization passes per lowering step (\optnum{}) to \{1, 3, 5, 7, 9\} and the number of compilations for differential testing (\diffnum{}) to \{2, 4, 6, 8, 10\}, respectively.
Notably, when examining the influence of one hyperparameter, we maintain the default setting for the other to ensure an isolated analysis.

\subsubsection{Metrics}
Following the existing work on MLIR testing~\cite{DBLP:conf/issta/Suo00JZW24, DBLP:conf/kbse/WangCXLWSZ23}, we used \textbf{the number of detected bugs} to measure the effectiveness of each studied technique.
For this metric, de-duplication is a necessary step~\cite{DBLP:journals/csur/ChenPPXZHZ20, DBLP:conf/issta/YangCFJS23, DBLP:conf/pldi/DonaldsonTTMMK21}.
In the study, for crash bugs, we de-duplicated them based on crash messages following the existing work~\cite{DBLP:conf/issta/Suo00JZW24, shen2024tale}.
For silent bugs, we de-duplicated them by analyzing their bug-triggering operations and passes, which are obtained based on delta debugging on both programs and passes~\cite{DBLP:conf/issta/YangCFJS23, DBLP:conf/pldi/DonaldsonTTMMK21}.
Then, we reported each bug to the developers for further confirmation.
Based on their feedback, the used de-duplication mechanisms are indeed accurate to a large extent.


\subsection{RQ1: Previously Unknown Bugs Detected by \tech{}}
\begin{table}[t]
\caption{Details of previously unknown bugs detected by \tech{}}
\label{tab:bug-detail}
\vspace{-3.5mm}
\resizebox{\columnwidth}{!}{
\footnotesize
\begin{threeparttable}
\begin{tabular}{l|lll||l|lll}
\toprule
Bug Id & Type         & Component               & Status    & Bug Id    & Type   & Component                   & Status    \\
\midrule
80960  & silent       & Documentation           & fixed     & 114652    & silent & Domain Specific(artih)      & duplicate \\
81228  & silent & Domain Specific(artih)  & fixed     & 114654    & silent & Domain Specific(artih)      & fixed     \\
82158  & silent       & Domain Specific(artih)  & fixed     & 114656    & silent & Domain Specific(linalg)     & confirmed \\
82168  & silent       & Domain Specific(artih)  & fixed     & 114657    & silent & Domain Specific(linalg)     & submitted \\
82622  & silent       & Domain Specific(math)   & submitted & 115293    & silent & Domain Specific(artih)      & fixed     \\
83530  & silent       & Domain Specific(artih)  & submitted & 115294\tnote{*} & silent & Domain Specific(vector)     & confirmed \\
92057  & crash        & General                 & fixed     & 115294\tnote{*} & crash  & Conversion                  & confirmed \\
94423  & crash        & Domain Specific(artih)  & fixed     & 116664    & silent & Domain Specific(scf) & submitted \\
94431  & silent       & Domain Specific(artih)  & fixed     & 118224    & crash  & Domain Specific(affine)     & submitted \\
95246  & crash        & Conversion              & fixed     & 118225    & crash  & Domain Specific(affine)     & submitted \\
102576 & crash        & Conversion              & fixed     & 126195    & silent & Domain Specific(artih)      & confirmed \\
102577 & crash        & Conversion              & submitted & 126196    & crash  & General                     & fixed     \\
111241 & crash        & Conversion               & submitted & 126197    & crash  & Domain Specific(vector)     & fixed     \\
111242 & crash        & Conversion              & fixed     & 126213    & crash  & General                     & confirmed \\
111243 & crash        & Domain Specific(linalg) & fixed     & 126371    & crash  & Domain Specific(vector)     & confirmed \\
111244 & crash        & Domain Specific(vector) & submitted & 128273    & silent & Domain Specific(affine)     & submitted \\
112878 & silent       & Domain Specific(linalg) & submitted & 128275    & crash  & Domain Specific(math)       & fixed     \\
112881 & silent       & Domain Specific(linalg) & fixed     & 128277    & crash  & Domain Specific(affine)     & fixed     \\
113687 & silent       & Domain Specific(affine) & submitted & 129414    & silent & General                     & submitted \\
113689 & silent       & General                 & submitted & 129415    & silent & Conversion                  & submitted \\
113690 & silent       & Domain Specific(linalg) & submitted & 129416    & crash  & General                     & confirmed \\
\bottomrule
\end{tabular}

\begin{tablenotes}
\footnotesize
\item[*] {\scriptsize These two bugs were reported together.}
\end{tablenotes}

\end{threeparttable}
}
\end{table}

Table~\ref{tab:bug-detail} provides the details of the previously unknown bugs detected by \tech{}, including the bug ID, type of bug, buggy component, and bug status. 
In total, \tech{} detected 42 bugs, comprising 23 silent bugs and 19 crash bugs. 
Among these, 12 
silent bugs 
and 14 
crash bugs have been confirmed or fixed by developers. 
However, existing techniques, such as MLIRSmith and MLIRod, cannot theoretically detect silent bugs due to the potential UB in their generated test programs and the absence of a lowering mechanism to compile these programs into executable forms.
The results underscore the \tech{} is effective in exposing silent bugs, which is orthogonal to all existing testing techniques for the MLIR compiler infrastructure.

\begin{figure}[t]
  \centering
\subfigure[Program snippet for triggering Bug\#80960 (silent) 
]{
    \label{fig:issue80960}
    \includegraphics[width=0.9\textwidth]{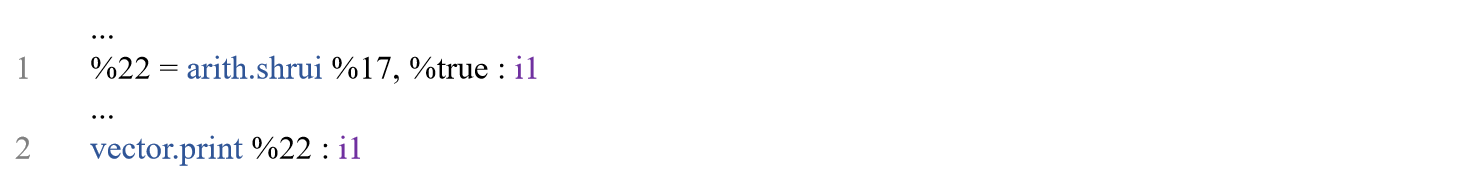}
  }
  \subfigure[Program snippet for triggering Bug\#81228 (silent)
  ]{
    \label{fig:issue81228}
    \includegraphics[width=0.9\textwidth]{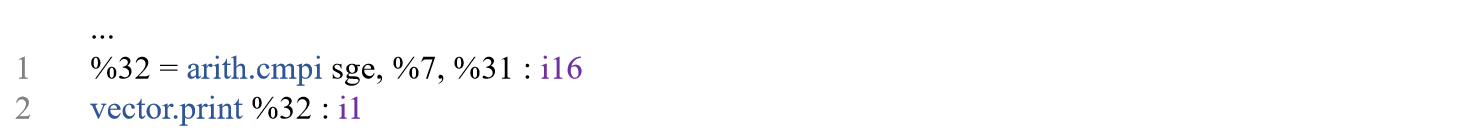}
  }
  \subfigure[Program snippet for triggering Bug\#102576 (crash)
  ]{
    \label{fig:issue102576}
    \includegraphics[width=0.9\textwidth]{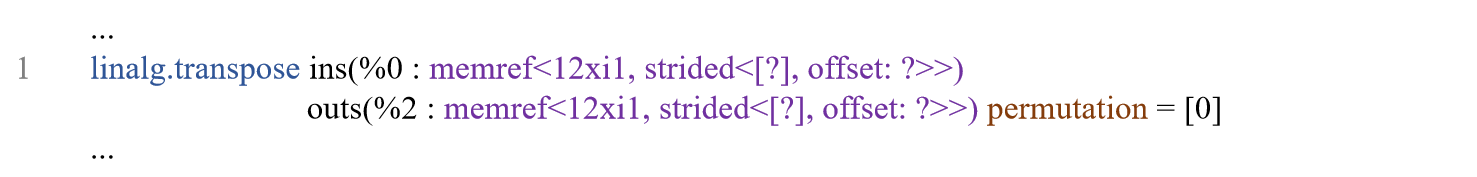}
  }
  \subfigure[Program snippet for triggering Bug\#126196 (crash)
  ]{
    \label{fig:issue126196}
    \includegraphics[width=0.9\textwidth]{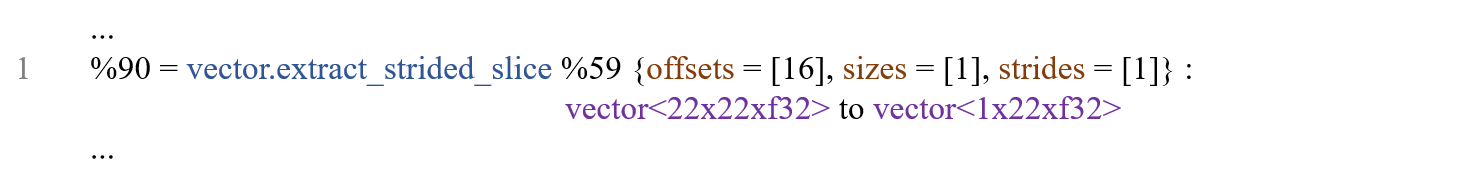}
  }
  \vspace{-4mm}
  \caption{previously unknown bug examples detected by \tech{}.}
  \label{fig:bugcase}
\end{figure}


From Table~\ref{tab:bug-detail}, we observed that the bugs detected by \tech{} span various components within the MLIR compiler infrastructure. These bugs are detailed as follows:

\textbf{Documentation} defines the semantics of operations.
Documentation changes can alter the semantics of MLIR operations behavior.
Bug in this category cannot be detected by existing approaches (e.g., MLIRSmith and MLIRod) since crashes can not reveal semantic issues in MLIR programs since they are not executed.
\tech{} identified a documentation-induced semantic bug (Figure~\ref{fig:issue80960}) manifested through inconsistent execution results. The core issue stems from conflicting specifications between three components: the LLVM support library function ({\tt APInt::lshr}), the LLVM IR instruction ({\tt lshr}), and the MLIR operation ({\tt arith.shrui}). While LLVM IR explicitly prohibits shift amounts equal to the bit width in {\tt lshr} (treating it as undefined behavior), both {\tt APInt::lshr} and the original {\tt arith.shrui} specification defined this edge case behavior. This discrepancy caused well-defined MLIR programs to exhibit undefined behavior when lowered to LLVM IR. The resolution involved aligning {\tt arith.shrui}'s semantics with LLVM IR by explicitly classifying this case as undefined. Notably, this latent bug persisted since {\tt arith.shrui}'s introduction and evaded detection by existing appraoches because it only manifested as behavioral inconsistencies rather than crashes.

\textbf{Domain-Specific Passes} are designed to apply specialized optimizations to MLIR programs, targeting specific types of operations or dialects that are relevant to a particular domain.
There are 28 bugs detected in these domain-specific passes by \tech{}, covering 6 dialects.
Specifically, there are 10 bugs in {\tt arith}-dialect specific passes, 2 bugs in {\tt math}-dialect specific passes, 6 bugs in {\tt linalg}-dialect specific passes, 4 bugs in {\tt vector}-dialect specific passes, 5 bugs in {\tt affine}-dialect specific passes and 1 bug in {\tt scf}-dialect specific passes.

Figure~\ref{fig:issue81228} shows an MLIR program that triggers a bug in this category.
The MLIR compiler generated different IRs for the given program across multiple runs under the same optimization due to the buggy data flow analysis in MLIR.
Specifically, the ``-int-range-optmizations'' pass utilizes a ``DataFlowSolver'' after performing fold optimizations.
However, when the fold optimization deletes an original operation and creates a new one at the same memory location, the solver fails to detect the change and returns the old operation's state. 
This bug is harmful as said by developers: ``its impact on other passes (e.g., Sparse Conditional Constant Propagation and dead code analysis) and makes debugging challenges''.
Developers have fixed it by adding a listener to track deleted operations, preventing the solver from returning outdated states.

\textbf{General passes} operate on MLIR programs at a more generic or broad level, typically affecting multiple dialects or operations. 
These passes are designed to be domain-agnostic, providing optimizations that are universally applicable rather than tailored to a specific use case. 6 bugs detected by \tech{} in this category.
Figure~\ref{fig:issue126196} illustrates an MLIR program that triggers a bug belonging to the general passes. 
The MLIR compiler crashed when executing the ``-canonicalize'' pass on the given MLIR program. 
Specifically, 
the {\tt vector.extract\_strided\_slice} operation functions as extracting a subvector {\tt \%90} from the source vector {\tt \%59}, where the beginning dimensions align with the {\tt sizes} attribute. 
When the {\tt sizes} attribute is simplified by specifying all its values as 1, with the remaining values inferred based on the shape of the source vector by the ``-canonicalize'' pass, the compiler will crash.
Since the ``-canonicalize'' pass lacks the necessary logic to handle it and led to an out-of-bounds access of the {\tt sizes} attribute during its inference, ultimately causing the crash.
To address this issue, the developers resolved the bug by adding bounds-checking logic to ensure safe access to the {\tt sizes} attribute.

\textbf{Conversion Passes} transform higher-level dialects into lower-level dialects. 
When conversion passes contain bugs, they either cannot produce executable IRs or generate erroneous ones.
7 bugs detected by \tech{} belong to this category.
Figure~\ref{fig:issue102576} illustrates an MLIR program that triggered a bug in the conversion pass (i.e., the ``-convert-linalg-to-loops'' pass).
Specifically, the pass incorrectly assumes that the \texttt{linalg.transpose} operation always produces a return value.
This operation works with both \texttt{memref} and \texttt{tensor} types, returning a tensor result for \texttt{tensor} input but no result for \texttt{memref} input.
The ``-convert-linalg-to-loops'' pass crashes when processing \texttt{linalg.transpose} on \texttt{memref} values because it incorrectly tries to read a non-existent return value.
Since \texttt{linalg.transpose} only returns a value for tensor operands, developers fixed the issue by adding type checks to skip return value handling for \texttt{memref} inputs.




These bugs are distributed across 4 categories with 9 
different components of the MLIR compiler infrastructure, demonstrating the effectiveness of \tech{} in bug detection. 

\subsection{RQ2: Compared to (Lifted) Existing MLIR Testing Techniques}
\label{sec:existing_tech_comparison}

\begin{table}[t!]
\caption{Comparison between \tech{} and lifted existing techniques in silent bug detection}
\label{tab:program}
\small
\vspace{-3.5mm}
\adjustbox{max width=\textwidth}{
\begin{tabular}{lccccc}
\toprule
Techniques    & \#Inconsistencies  & \#FP Inconsistencies   & FP Rate & \#TP Inconsistencies & \#Silent Bugs\\
\midrule
MLIRSmith$_{\textit{enhanced}}$ & 4,914  & 4,783 &  97.33\% & 131 & 14 \\
MLIRod$_{\textit{enhanced}}$    & 4,542  & 4,404 & 96.96\%  & 138 & 15 \\
\tech{}$_{\textit{smith}}$      & 519    & 0     & 0\%      & 519   & 25 \\
\tech{}$_{\textit{od}}$         & 470    & 0     & 0\%      & 470   & 31 \\
\bottomrule
\end{tabular}
}
\end{table}

As presented in Section~\ref{sec:baseline}, we lifted both MLIRSmith and MLIRod as MLIRSmith$_{\textit{enhanced}}$ and MLIRod$_{\textit{enhanced}}$, enabling the comparison between \tech{} and the existing MLIR testing techniques in silent bug detection.
Table~\ref{tab:program} presents the comparison results among these studied techniques in silent MLIR bug detection during the given testing time.
In this table, Columns 2-6 represent the number of inconsistencies detected via differential testing, the number of false positives among these inconsistencies, the ratio of false positives (dividing the number of false positives by the total number of inconsistencies), the number of true inconsistencies caused by silent bugs, the number of silent bugs after de-duplicating inconsistencies, respectively.
For each detected inconsistency by MLIRSmith$_{\textit{enhanced}}$ or MLIRod$_{\textit{enhanced}}$, we applied \tech{} to check whether the corresponding MLIR program has UB and then eliminate it.
If the inconsistency still exists by running the UB-free program, it is regarded as a true inconsistency; Otherwise, it is a false positive.
Indeed, through our manual analysis on these true inconsistencies detected by existing techniques and the inconsistencies detected by \tech{}, all of them are real bugs.


From Table~\ref{tab:program}, \tech{}$_{\textit{smith}}$ (25) and \tech{}$_{\textit{od}}$ (31) detected more silent bugs than MLIRSmith$_{\textit{enhanced}}$ (14) or MLIRod$_{\textit{enhanced}}$ (15), respectively, demonstrating the effectiveness of \tech{} in detecting silent MLIR bugs.
Although MLIRSmith$_{\textit{enhanced}}$ and MLIRod$_{\textit{enhanced}}$ were able to detect some silent bugs, they suffered from extremely high false positive rates (97.33\% and 96.96\%, respectively).
Moreover, \tech{} played a crucial role in distinguishing the silent bugs detected by MLIRSmith${_\textit{enhanced}}$ and MLIRod${_\textit{enhanced}}$ from a large number of false positive inconsistencies.
This indicates that even with enhancements, MLIRSmith and MLIRod remain impractical for reliably detecting silent bugs.



\begin{figure}[t]
  \centering
  

  \subfigure[\tech{}$_{\textit{smith}}$ v.s MLIRSmith$_{\textit{enhanced}}$ 
  ]{
    \label{fig:rq2:1}
    \includegraphics[width=0.3\textwidth]{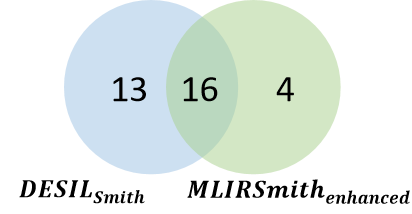}
  }
  \subfigure[\tech{}$_{\textit{od}}$ v.s. MLIRod$_{\textit{enhanced}}$ 
  ]{
    \label{fig:rq2:2}
    \includegraphics[width=0.3\textwidth]{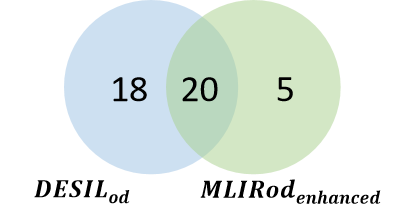}
  }

    \vspace{-4mm}
  \caption{The number of unique bugs detected by \tech{} and lifted existing techniques in bug detection.}
  \vspace{-4mm}
  \label{fig:rq2}
\end{figure}

Besides silent bugs, all these studied techniques are able to detect crash bugs. 
Hence, we further compared them in the overall bug detection capability.
In total, \tech{}$_{\textit{smith}}$ and \tech{}$_{\textit{od}}$ detected 29 and 38 MLIR bugs respectively, while MLIRSmith$_{\textit{enhanced}}$ and MLIRod$_{\textit{enhanced}}$ detected 20 and 25 bugs respectively, demonstrating the superiority of \tech{} over baselines in terms of overall bug detection effectiveness.
Figure~\ref{fig:rq2} shows the overlap analysis results for \tech{}$_{\textit{smith}}$ v.s MLIRSmith$_{\textit{enhanced}}$ and \tech{}$_{\textit{od}}$ v.s. MLIRod$_{\textit{enhanced}}$.
From this figure, by comparing \tech{}$_{\textit{smith}}$ with MLIRSmith$_{\textit{enhanced}}$, the former detected 13 unique bugs (including 12 silent bugs and one crash bug) while the latter detected only 4 unique bugs (including one silent bugs and 3 crash bugs).
Similarly, by comparing \tech{}$_{\textit{od}}$ with MLIRod$_{\textit{enhanced}}$, the former detected 18 unique bugs (including 16 silent bugs and 2 crash bugs) while the latter detected only 5 unique bugs (including 0 silent bug and 5 crash bugs).
The results further confirm the effectiveness of \tech{}.
Through further observation, we found that \tech{}$_{\textit{smith}}$ (4) and \tech{}$_{\textit{od}}$ (7) detected slightly less crash bugs than MLIRSmith$_{\textit{enhanced}}$ (6) and MLIRod$_{\textit{enhanced}}$ (10), respectively.
This is as expected, since \tech{} requires extra time cost for UB elimination, and thus ran less MLIR programs for testing.
Specifically, during the same testing period, the number of executed programs for \tech{}$_{\textit{smith}}$ and \tech{}$_{\textit{od}}$ is 13,269 and 13,541 respectively, while that for MLIRSmith$_{\textit{enhanced}}$ and MLIRod$_{\textit{enhanced}}$ is 21,681 and 21,420 respectively.
Nevertheless, the strong capability of \tech{} in detecting silent bugs far outweighs its slight drawback in crash bug detection, which is due to the additional time cost incurred by UB elimination.

\subsection{RQ3: Ablation Study}
\label{sec:component}
We first investigated the contribution of the lowering path optimization mechanism in \tech{} by running \tech{}$_{\textit{smith}}^{\textit{w/o lower}}$ and \tech{}$_{\textit{od}}^{\textit{w/o lower}}$.
Over five repeated 12-hour fuzzing sessions, both \tech{}$_{\textit{smith}}^{\textit{w/o lower}}$ and \tech{}$_{\textit{od}}^{\textit{w/o lower}}$ failed to compile any MLIR program into an executable form, even when applying 50 lowering passes — more than twice the number typically required by \tech{} (i.e., 21 on average).
That is, randomly selecting a sequence of lowering passes hardly succeeds in converting all operations from diverse dialects into an executable dialect within a reasonable number of steps.
This underscores the critical role of our lowering path optimization strategy in ensuring the feasibility of \tech{} for silent bug detection.

\begin{figure}[t]
  \centering
  \subfigure[$\tech{}_{smith}$ and its variant]{
    \label{fig:woopt:smith}
    \includegraphics[width=0.24\textwidth]{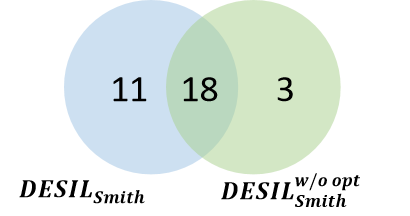}
  }\hspace{8mm}
  \subfigure[$\tech{}_{od}$ and its variant]{
    \label{fig:woopt:od}
    \includegraphics[width=0.24\textwidth]{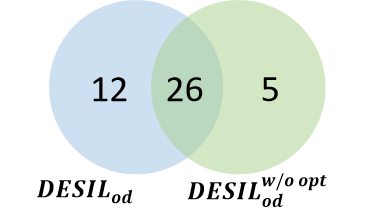}
  }
  \vspace{-4mm}
  \caption{The number of unique bugs with and without optimization recommendation component.}
  \label{fig:woopt}
\end{figure}

We then investigated the contribution of the operation-aware optimization specifying strategy in \tech{} by comparing with \tech{}$_{\textit{smith}}^{\textit{w/o opt}}$ and \tech{}$_{\textit{od}}^{\textit{w/o opt}}$.
During the given testing testing, \tech{}$_{\textit{smith}}$ and \tech{}$_{\textit{od}}$ detected 29 and 38 bugs while \tech{}$_{\textit{smith}}^{\textit{w/o opt}}$ and \tech{}$_{\textit{od}}^{\textit{w/o opt}}$ detected 21 and 31 bugs respectively.
Figure~\ref{fig:woopt} further shows the overlap analysis results among the studied techniques.
As shown in the figure, \tech{}$_{\textit{smith}}$ detected 11 unique bugs compared to \tech{}$_{\textit{smith}}^{\textit{w/o opt}}$, and \tech{}$_{\textit{od}}$ detected 12 unique bugs compared to \tech{}$_{\textit{od}}^{\textit{w/o opt}}$.
In contrast, \tech{}$_{\textit{smith}}^{\textit{w/o opt}}$ and \tech{}$_{\textit{od}}^{\textit{w/o opt}}$ detected only 3 and 5 unique bugs, respectively.
These results highlight the superiority of the operation-aware optimization specifying strategy in \tech{} over the random strategy for specifying optimization passes.

\subsection{RQ4: Influence of different configurations}
\label{sec:param}

\begin{figure}[t]
\centering
\subfigure[The number of detected bugs with different \optnum{}
]{
  \label{fig:optnum}
  \includegraphics[width=0.490374\columnwidth]{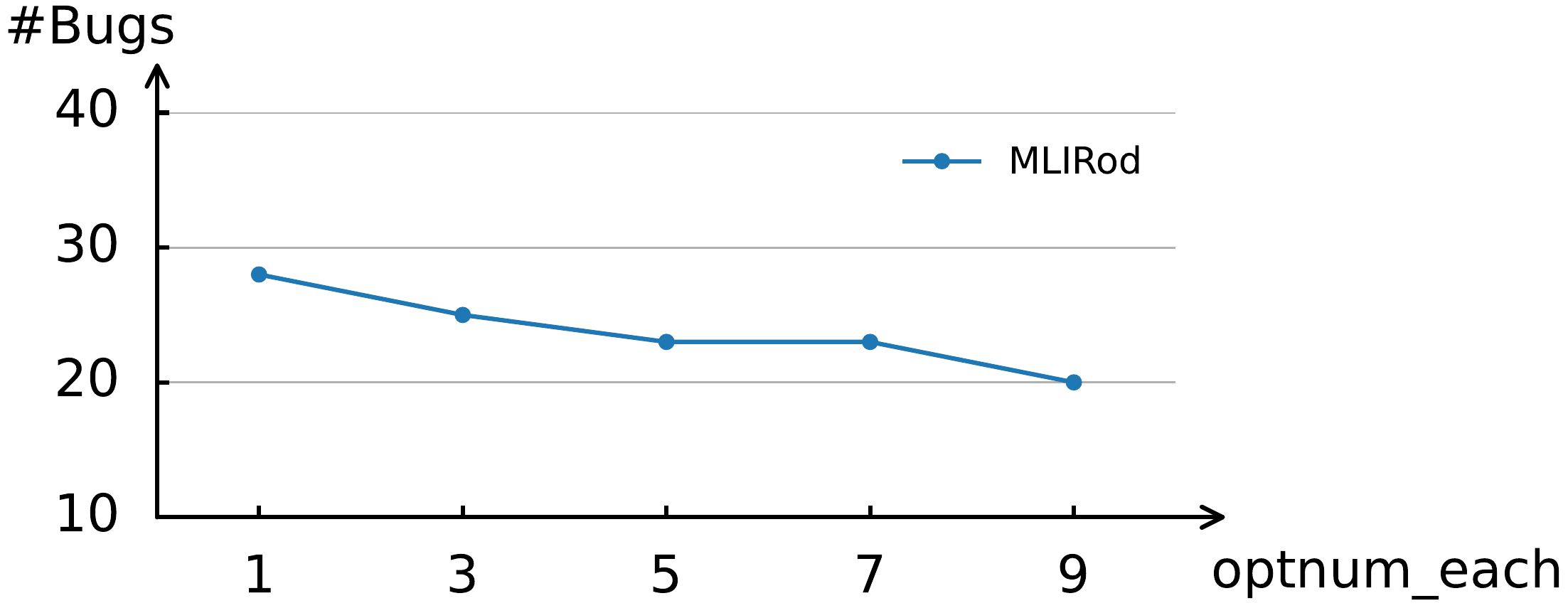}
}
\subfigure[The number of detected bugs with different \diffnum{}
]{
  \label{fig:diffnum}
  \includegraphics[width=0.437\columnwidth]{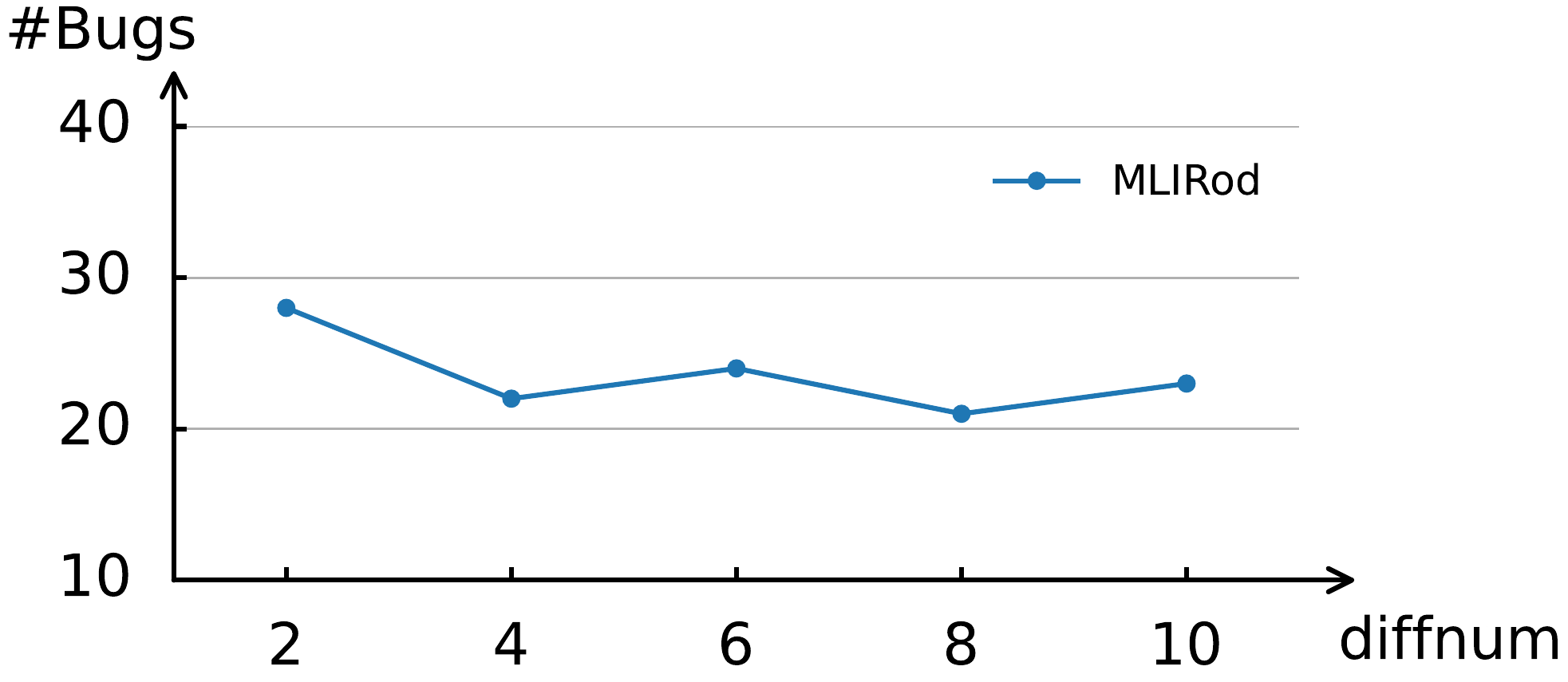}
}
\vspace{-4mm}
\caption{The number of bugs detected by \tech{} under different configurations. }
\label{fig:param}
\end{figure}


Figure~\ref{fig:param} shows the effectiveness of \tech{} under different configurations of \optnum{} (the number of optimization passes per lowering step) and \diffnum{} (the number of compilations for differential testing), respectively.
The y-axis represents the total number of bugs detected by \tech{}$_{\textit{smith}}$ and \tech{}$_{\textit{od}}$.
In general, as the values of \optnum{} or \diffnum{} increase, the effectiveness of \tech{} decreases to some extent (especially for the former).
This phenomenon arises from the trade-off between bug detection capability and time efficiency.
While larger values enable broader exploration of the optimization space, potentially increasing the likelihood of uncovering bugs, the associated time overhead ultimately reduces these benefits.
This aligns with the conclusion of an existing study~\cite{DBLP:conf/icse/ChenHHXZ0X16}, which highlights testing efficiency as one of the most critical factors in compiler testing.
Similarly, the decreasing trend for \diffnum{} is less pronounced, as the overhead it incurs under these settings is lower than that of \optnum{}.
Based on this insight, we set \optnum{} to 1 and \diffnum{} to 2 as the default configurations of \tech{} for practical use.


\section{Discussion}
\noindent \textbf{Significance of \tech{}.}
While \tech{} is designed to fuzz the MLIR compiler infrastructure, its impact extends beyond a single system. Many compilers, such as Flang~\cite{Flang} and IREE~\cite{IREE}, are built on top of MLIR, meaning that improving the reliability of the MLIR compiler infrastructure enhances the quality and robustness of all compilers that depend on it. In other words, fuzzing the MLIR compiler infrastructure has a far-reaching effect, benefiting multiple compiler systems rather than just one.

Moreover, \tech{} is independent of MLIR test program generation techniques and can be integrated with any of them.
Specifically, for any given MLIR program, \tech{} can transform it into a UB-free version and then compile it into an executable form for silent bug detection.
Our evaluation has demonstrated the effectiveness of \tech{} when combined with two state-of-the-art MLIR program generation tools, i.e., MLIRSmith and MLIRod.
Therefore, we are confident that \tech{} can also enhance future, more advanced MLIR program generation techniques, further extending its impact and applicability.

\smallskip
\noindent\textbf{Extension of \tech{}.}
Currently, 
\tech{} supports most of the widely used dialects and operations in MLIR fuzzing, 
specifically the dialects and operations supported by MLIRSmith. 
However, the dialects supported by MLIRSmith are primarily middle-level dialects, meaning that certain higher-level dialects, such as the TOSA dialect, are not yet supported. 
Fortunately, \tech{} has already defined a comprehensive set of undefined behavior elimination rules, 
which can be reused to handle undefined behaviors in the TOSA dialect. 
As a result, supporting new dialects in \tech{} generally requires only a minimal number of additional elimination rules.
Another area where \tech{} can be improved is in expanding the range of supported execution platforms. 
To achieve this, 
adding support for a new executable dialect can be accomplished by adjusting the operation-specific lowering path within \tech{}. 
No additional mechanisms are required to enable the support of new executable dialects.
For the post-processing of bug reports, 
we currently reduce MLIR programs manually. 
This involves automatically localizing problematic operations and manually reducing the program using define-use chains. 
In the future, tools like Creduce could be introduced to automate and streamline the program reduction process.

\smallskip
\noindent\textbf{Threats to Validity.}
The threat to \textit{internal} validity primarily concerns the implementation of \tech{}. To mitigate this, two authors meticulously reviewed the source code and designed unit tests to ensure the correctness of \tech{}. 
In addition, we further validated \tech{} by applying the ``-generate-runtime-verification'' pass, which is equipped in the MLIR compiler for verifing the correctness of operations, to 10,000 test programs generated by \tech{}. 
None of the test programs triggered the check failure in the ``generate-runtime-verification'' pass, demonstrating the stability of \tech{}.
For the existing techniques (i.e., MLIRSmith and MLIRod), we directly adopted their publicly released implementations and followed the recommended settings.
The threat to \textit{external} validity arises from the choice of the subject under test. To address this, we selected the latest versions of the MLIR compiler infrastructure as the subject and conducted continuous fuzzing to thoroughly evaluate the effectiveness of \tech{} in detecting previously unknown bugs.
The threat to \textit{construct} validity arises from the randomness in the experiments and the hyper-parameter settings in \tech{}.
To mitigate the impact of randomness, we repeated each experiment five times (three times for parameter-setting experiments due to their extensive time cost).
To address concerns regarding hyper-parameter settings, we evaluated \tech{} under various configurations, as detailed in Section~\ref{sec:param}.

\section{Related Work}

In recent years, several testing techniques have been proposed for the MLIR compiler infrastructure~\cite{DBLP:conf/issta/Suo00JZW24, DBLP:conf/kbse/WangCXLWSZ23}.
For example, Wang et al. introduced MLIRSmith~\cite{DBLP:conf/kbse/WangCXLWSZ23}, the first MLIR program generator designed for testing the MLIR compiler infrastructure.
MLIRSmith generates MLIR programs by first constructing program templates based on MLIR's grammar and then filling these templates according to semantic rules to ensure the generation of semantically valid MLIR programs.
Suo et al. proposed MLIRod~\cite{DBLP:conf/issta/Suo00JZW24}, which defines operation dependency coverage as the testing guidance and employs four types of dependency-specific mutations to generate new MLIR programs, enhancing the effectiveness of testing the MLIR compiler infrastructure.
They have demonstrated significant effectiveness in detecting crash bugs by generating semantically valid MLIR programs.

As explained in Section~\ref{sec:intro}, all existing techniques suffer from the UB issue, which prevents them from detecting silent bugs.
Specifically, the MLIR programs generated by these techniques are highly likely to exhibit UB, leading to unpredictable execution results and consequently numerous false positives, as confirmed by our experiment shown in Section~\ref{sec:existing_tech_comparison}.
Due to this issue, none of these techniques incorporate a lowering process to compile MLIR programs into executable programs for execution, making it impossible for them to detect silent bugs.
In contrast, our work introduces \tech{}, the first technique designed to detect silent MLIR bugs by addressing the UB issue through a set of UB-elimination rules and designing a lowering path optimization strategy for checking program execution.
Thus, \tech{} is orthogonal to all existing MLIR testing techniques, offering a significant improvement in quality assurance for the MLIR compiler infrastructure.


Additionally, there are a lot of testing techniques for traditional compilers in the literature~\cite{DBLP:conf/pldi/YangCER11, DBLP:journals/pacmpl/LivinskiiBR20, DBLP:conf/issta/SharmaYD23, DBLP:journals/pacmpl/LecoeurMD23}. 
Some of these techniques are capable of detecting silent bugs in traditional compilers by ensuring UB-free test programs.
For example, Csmith~\cite{DBLP:conf/pldi/YangCER11}, a grammar-based C program generator, leverages predefined rules (such as safe arithmetic wrappers) and built-in dynamic checks to prevent undefined behaviors during test program generation. 
Yarpgen~\cite{DBLP:journals/pacmpl/LivinskiiBR20}, another C/C++ program generator, ensures expression safety by defining a set of safe expressions that prevent undefined behaviors at the generation phase.
RustSmith~\cite{DBLP:conf/issta/SharmaYD23} maintains a symbol table and enforces safety rules to validate borrowing rules and variable lifetimes, ensuring the generation of well-defined Rust programs.
Lecoeur et al. introduced reconditioning~\cite{DBLP:journals/pacmpl/LecoeurMD23}, a technique that eliminates undefined behaviors in OpenGL Shading Language (GLSL) and WebGPU Shading Language (WGSL) through post-processing with program transformations.

Reconditioning is the most relevant technique to \tech{}, as both address UB through post-processing rather than during program generation and rely on code transformations for UB elimination.
However, \tech{} differs in several key aspects.
First, \tech{} targets a fundamentally different domain, focusing on intermediate representations (IRs) rather than high-level languages.
Reconditioning operates on languages with restricted memory allocation (e.g., GLSL, which disallows variable-length arrays), while \tech{} handles IRs that support flexible memory allocation (e.g., dynamic shapes) and complex operation semantics (e.g., {\tt linalg.matmul} for matrix multiplication).
As a result, \tech{} must address unique categories of UB and requires more sophisticated transformation rules.
Second, \tech{} introduces an additional challenge absent in reconditioning: lowering path optimization.
MLIR compilation involves multiple dialects, requiring careful selection of lowering passes to ensure successful translation to an executable representation.
\tech{} tackles this problem with a structured lowering-path optimization strategy, making it fundamentally distinct from reconditioning.


\section{Conclusion}

In this paper, we presented \tech{}, a novel technique designed to bridge the gap in detecting silent bugs in the MLIR compiler infrastructure by jointly generating UB-free programs and determining optimal lowering paths. 
\tech{} addresses two key challenges in MLIR bug detection: (1) eliminating undefined behavior (UB) in UB-prone operations through a set of undefined behavior elimination rules, and (2) determining an optimal lowering path to prevent redundant or circular application of lowering passes, ensuring efficient compilation to an executable representation. 
By incorporating a differential testing oracle, 
\tech{} further enhances its ability to detect silent bugs by comparing the results of executable programs affected by different optimization passes.

Our evaluation demonstrates \tech{}'s effectiveness in detecting silent MLIR bugs, identifying 42 previously unknown bugs (23 silent and 19 crash bugs) over a four-month testing period, with 18 fixed and 26 confirmed by the developers. 
Exhaustive experiments highlight the critical contributions of \tech{}'s UB elimination, lowering path optimization, and optimization recommendation mechanism, showcasing its ability to significantly improve bug detection accuracy and efficiency compared to baseline approaches.

\section{Data Availability}

We released the source code of \tech{} (implemented in 4.6K lines of C++ code), along with all experimental data at our project homepage ~\cite{DESILrepository}.


\bibliographystyle{ACM-Reference-Format}
\bibliography{sample-base}


\begin{thebibliography}{21}


\ifx \showCODEN    \undefined \def \showCODEN     #1{\unskip}     \fi
\ifx \showDOI      \undefined \def \showDOI       #1{#1}\fi
\ifx \showISBNx    \undefined \def \showISBNx     #1{\unskip}     \fi
\ifx \showISBNxiii \undefined \def \showISBNxiii  #1{\unskip}     \fi
\ifx \showISSN     \undefined \def \showISSN      #1{\unskip}     \fi
\ifx \showLCCN     \undefined \def \showLCCN      #1{\unskip}     \fi
\ifx \shownote     \undefined \def \shownote      #1{#1}          \fi
\ifx \showarticletitle \undefined \def \showarticletitle #1{#1}   \fi
\ifx \showURL      \undefined \def \showURL       {\relax}        \fi
\providecommand\bibfield[2]{#2}
\providecommand\bibinfo[2]{#2}
\providecommand\natexlab[1]{#1}
\providecommand\showeprint[2][]{arXiv:#2}

\bibitem[aff(2025)]%
        {affine}
 \bibinfo{year}{2025}\natexlab{}.
\newblock \bibinfo{title}{{Affine Documentation}}.
\newblock
\newblock
\newblock
\shownote{{https://mlir.llvm.org/docs/Dialects/Affine}}.


\bibitem[DES(2025)]%
        {DESILrepository}
 \bibinfo{year}{2025}\natexlab{}.
\newblock \bibinfo{title}{{DESIL repository}}.
\newblock
\newblock
\newblock
\shownote{{https://github.com/DESIL-tech/DESIL}}.


\bibitem[Fla(2025)]%
        {Flang}
 \bibinfo{year}{2025}\natexlab{}.
\newblock \bibinfo{title}{{Flang}}.
\newblock
\newblock
\newblock
\shownote{{https://github.com/llvm/llvm-project/tree/main/flang}}.


\bibitem[IRE(2025)]%
        {IREE}
 \bibinfo{year}{2025}\natexlab{}.
\newblock \bibinfo{title}{{IREE}}.
\newblock
\newblock
\newblock
\shownote{{https://github.com/iree-org/iree}}.


\bibitem[MLI(2025)]%
        {MLIR_language_reference}
 \bibinfo{year}{2025}\natexlab{}.
\newblock \bibinfo{title}{{MLIR language reference}}.
\newblock
\newblock
\newblock
\shownote{{https://mlir.llvm.org/docs/LangRef}}.


\bibitem[Aho et~al\mbox{.}(1986)]%
        {DBLP:books/aw/AhoSU86}
\bibfield{author}{\bibinfo{person}{Alfred~V. Aho}, \bibinfo{person}{Ravi Sethi}, {and} \bibinfo{person}{Jeffrey~D. Ullman}.} \bibinfo{year}{1986}\natexlab{}.
\newblock \bibinfo{booktitle}{\emph{Compilers: Principles, Techniques, and Tools}}.
\newblock \bibinfo{publisher}{Addison-Wesley}.
\newblock
\showISBNx{0-201-10088-6}
\urldef\tempurl%
\url{https://www.worldcat.org/oclc/12285707}
\showURL{%
\tempurl}


\bibitem[Chen et~al\mbox{.}(2016)]%
        {DBLP:conf/icse/ChenHHXZ0X16}
\bibfield{author}{\bibinfo{person}{Junjie Chen}, \bibinfo{person}{Wenxiang Hu}, \bibinfo{person}{Dan Hao}, \bibinfo{person}{Yingfei Xiong}, \bibinfo{person}{Hongyu Zhang}, \bibinfo{person}{Lu Zhang}, {and} \bibinfo{person}{Bing Xie}.} \bibinfo{year}{2016}\natexlab{}.
\newblock \showarticletitle{An empirical comparison of compiler testing techniques}. In \bibinfo{booktitle}{\emph{Proceedings of the 38th International Conference on Software Engineering, {ICSE} 2016, Austin, TX, USA, May 14-22, 2016}}, \bibfield{editor}{\bibinfo{person}{Laura~K. Dillon}, \bibinfo{person}{Willem Visser}, {and} \bibinfo{person}{Laurie~A. Williams}} (Eds.). \bibinfo{publisher}{{ACM}}, \bibinfo{pages}{180--190}.
\newblock
\urldef\tempurl%
\url{https://doi.org/10.1145/2884781.2884878}
\showDOI{\tempurl}


\bibitem[Chen et~al\mbox{.}(2021)]%
        {DBLP:journals/csur/ChenPPXZHZ20}
\bibfield{author}{\bibinfo{person}{Junjie Chen}, \bibinfo{person}{Jibesh Patra}, \bibinfo{person}{Michael Pradel}, \bibinfo{person}{Yingfei Xiong}, \bibinfo{person}{Hongyu Zhang}, \bibinfo{person}{Dan Hao}, {and} \bibinfo{person}{Lu Zhang}.} \bibinfo{year}{2021}\natexlab{}.
\newblock \showarticletitle{A Survey of Compiler Testing}.
\newblock \bibinfo{journal}{\emph{{ACM} Comput. Surv.}} \bibinfo{volume}{53}, \bibinfo{number}{1} (\bibinfo{year}{2021}), \bibinfo{pages}{4:1--4:36}.
\newblock
\urldef\tempurl%
\url{https://doi.org/10.1145/3363562}
\showDOI{\tempurl}


\bibitem[Donaldson et~al\mbox{.}(2021)]%
        {DBLP:conf/pldi/DonaldsonTTMMK21}
\bibfield{author}{\bibinfo{person}{Alastair~F. Donaldson}, \bibinfo{person}{Paul Thomson}, \bibinfo{person}{Vasyl Teliman}, \bibinfo{person}{Stefano Milizia}, \bibinfo{person}{Andr{\'{e}}~Perez Maselco}, {and} \bibinfo{person}{Antoni Karpinski}.} \bibinfo{year}{2021}\natexlab{}.
\newblock \showarticletitle{Test-case reduction and deduplication almost for free with transformation-based compiler testing}. In \bibinfo{booktitle}{\emph{{PLDI} '21: 42nd {ACM} {SIGPLAN} International Conference on Programming Language Design and Implementation, Virtual Event, Canada, June 20-25, 2021}}, \bibfield{editor}{\bibinfo{person}{Stephen~N. Freund} {and} \bibinfo{person}{Eran Yahav}} (Eds.). \bibinfo{publisher}{{ACM}}, \bibinfo{pages}{1017--1032}.
\newblock
\urldef\tempurl%
\url{https://doi.org/10.1145/3453483.3454092}
\showDOI{\tempurl}


\bibitem[Harrold and Soffa(1994)]%
        {DBLP:journals/toplas/HarroldS94}
\bibfield{author}{\bibinfo{person}{Mary~Jean Harrold} {and} \bibinfo{person}{Mary~Lou Soffa}.} \bibinfo{year}{1994}\natexlab{}.
\newblock \showarticletitle{Efficient Computation of Interprocedural Definition-Use Chains}.
\newblock \bibinfo{journal}{\emph{{ACM} Trans. Program. Lang. Syst.}} \bibinfo{volume}{16}, \bibinfo{number}{2} (\bibinfo{year}{1994}), \bibinfo{pages}{175--204}.
\newblock
\urldef\tempurl%
\url{https://doi.org/10.1145/174662.174663}
\showDOI{\tempurl}


\bibitem[Lattner et~al\mbox{.}(2021)]%
        {mlir}
\bibfield{author}{\bibinfo{person}{Chris Lattner}, \bibinfo{person}{Mehdi Amini}, \bibinfo{person}{Uday Bondhugula}, \bibinfo{person}{Albert Cohen}, \bibinfo{person}{Andy Davis}, \bibinfo{person}{Jacques Pienaar}, \bibinfo{person}{River Riddle}, \bibinfo{person}{Tatiana Shpeisman}, \bibinfo{person}{Nicolas Vasilache}, {and} \bibinfo{person}{Oleksandr Zinenko}.} \bibinfo{year}{2021}\natexlab{}.
\newblock \showarticletitle{{{MLIR}}: Scaling Compiler Infrastructure for Domain Specific Computation}. In \bibinfo{booktitle}{\emph{2021 {{IEEE/ACM}} International Symposium on Code Generation and Optimization (CGO)}}. \bibinfo{pages}{2--14}.
\newblock
\urldef\tempurl%
\url{https://doi.org/10.1109/CGO51591.2021.9370308}
\showDOI{\tempurl}


\bibitem[Lecoeur et~al\mbox{.}(2023)]%
        {DBLP:journals/pacmpl/LecoeurMD23}
\bibfield{author}{\bibinfo{person}{Bastien Lecoeur}, \bibinfo{person}{Hasan Mohsin}, {and} \bibinfo{person}{Alastair~F. Donaldson}.} \bibinfo{year}{2023}\natexlab{}.
\newblock \showarticletitle{Program Reconditioning: Avoiding Undefined Behaviour When Finding and Reducing Compiler Bugs}.
\newblock \bibinfo{journal}{\emph{Proc. {ACM} Program. Lang.}} \bibinfo{volume}{7}, \bibinfo{number}{{PLDI}} (\bibinfo{year}{2023}), \bibinfo{pages}{1801--1825}.
\newblock
\urldef\tempurl%
\url{https://doi.org/10.1145/3591294}
\showDOI{\tempurl}


\bibitem[Livinskii et~al\mbox{.}(2020)]%
        {DBLP:journals/pacmpl/LivinskiiBR20}
\bibfield{author}{\bibinfo{person}{Vsevolod Livinskii}, \bibinfo{person}{Dmitry Babokin}, {and} \bibinfo{person}{John Regehr}.} \bibinfo{year}{2020}\natexlab{}.
\newblock \showarticletitle{Random testing for {C} and {C++} compilers with YARPGen}.
\newblock \bibinfo{journal}{\emph{Proc. {ACM} Program. Lang.}} \bibinfo{volume}{4}, \bibinfo{number}{{OOPSLA}} (\bibinfo{year}{2020}), \bibinfo{pages}{196:1--196:25}.
\newblock
\urldef\tempurl%
\url{https://doi.org/10.1145/3428264}
\showDOI{\tempurl}


\bibitem[Livinskii et~al\mbox{.}(2023)]%
        {DBLP:journals/pacmpl/LivinskiiBR23}
\bibfield{author}{\bibinfo{person}{Vsevolod Livinskii}, \bibinfo{person}{Dmitry Babokin}, {and} \bibinfo{person}{John Regehr}.} \bibinfo{year}{2023}\natexlab{}.
\newblock \showarticletitle{Fuzzing Loop Optimizations in Compilers for {C++} and Data-Parallel Languages}.
\newblock \bibinfo{journal}{\emph{Proc. {ACM} Program. Lang.}} \bibinfo{volume}{7}, \bibinfo{number}{{PLDI}} (\bibinfo{year}{2023}), \bibinfo{pages}{1826--1847}.
\newblock
\urldef\tempurl%
\url{https://doi.org/10.1145/3591295}
\showDOI{\tempurl}


\bibitem[Sharma et~al\mbox{.}(2023)]%
        {DBLP:conf/issta/SharmaYD23}
\bibfield{author}{\bibinfo{person}{Mayank Sharma}, \bibinfo{person}{Pingshi Yu}, {and} \bibinfo{person}{Alastair~F. Donaldson}.} \bibinfo{year}{2023}\natexlab{}.
\newblock \showarticletitle{RustSmith: Random Differential Compiler Testing for Rust}. In \bibinfo{booktitle}{\emph{Proceedings of the 32nd {ACM} {SIGSOFT} International Symposium on Software Testing and Analysis, {ISSTA} 2023, Seattle, WA, USA, July 17-21, 2023}}, \bibfield{editor}{\bibinfo{person}{Ren{\'{e}} Just} {and} \bibinfo{person}{Gordon Fraser}} (Eds.). \bibinfo{publisher}{{ACM}}, \bibinfo{pages}{1483--1486}.
\newblock
\urldef\tempurl%
\url{https://doi.org/10.1145/3597926.3604919}
\showDOI{\tempurl}


\bibitem[Shen et~al\mbox{.}(2024)]%
        {shen2024tale}
\bibfield{author}{\bibinfo{person}{Qingchao Shen}, \bibinfo{person}{Yongqiang Tian}, \bibinfo{person}{Haoyang Ma}, \bibinfo{person}{Junjie Chen}, \bibinfo{person}{Lili Huang}, \bibinfo{person}{Ruifeng Fu}, \bibinfo{person}{Shing-Chi Cheung}, {and} \bibinfo{person}{Zan Wang}.} \bibinfo{year}{2024}\natexlab{}.
\newblock \showarticletitle{A Tale of Two DL Cities: When Library Tests Meet Compiler}.
\newblock \bibinfo{journal}{\emph{arXiv preprint arXiv:2407.16626}} (\bibinfo{year}{2024}).
\newblock


\bibitem[Suo et~al\mbox{.}(2024)]%
        {DBLP:conf/issta/Suo00JZW24}
\bibfield{author}{\bibinfo{person}{Chenyao Suo}, \bibinfo{person}{Junjie Chen}, \bibinfo{person}{Shuang Liu}, \bibinfo{person}{Jiajun Jiang}, \bibinfo{person}{Yingquan Zhao}, {and} \bibinfo{person}{Jianrong Wang}.} \bibinfo{year}{2024}\natexlab{}.
\newblock \showarticletitle{Fuzzing {MLIR} Compiler Infrastructure via Operation Dependency Analysis}. In \bibinfo{booktitle}{\emph{Proceedings of the 33rd {ACM} {SIGSOFT} International Symposium on Software Testing and Analysis, {ISSTA} 2024, Vienna, Austria, September 16-20, 2024}}, \bibfield{editor}{\bibinfo{person}{Maria Christakis} {and} \bibinfo{person}{Michael Pradel}} (Eds.). \bibinfo{publisher}{{ACM}}, \bibinfo{pages}{1287--1299}.
\newblock
\urldef\tempurl%
\url{https://doi.org/10.1145/3650212.3680360}
\showDOI{\tempurl}


\bibitem[Wang et~al\mbox{.}(2023)]%
        {DBLP:conf/kbse/WangCXLWSZ23}
\bibfield{author}{\bibinfo{person}{Haoyu Wang}, \bibinfo{person}{Junjie Chen}, \bibinfo{person}{Chuyue Xie}, \bibinfo{person}{Shuang Liu}, \bibinfo{person}{Zan Wang}, \bibinfo{person}{Qingchao Shen}, {and} \bibinfo{person}{Yingquan Zhao}.} \bibinfo{year}{2023}\natexlab{}.
\newblock \showarticletitle{MLIRSmith: Random Program Generation for Fuzzing {MLIR} Compiler Infrastructure}. In \bibinfo{booktitle}{\emph{38th {IEEE/ACM} International Conference on Automated Software Engineering, {ASE} 2023, Luxembourg, September 11-15, 2023}}. \bibinfo{publisher}{{IEEE}}, \bibinfo{pages}{1555--1566}.
\newblock
\urldef\tempurl%
\url{https://doi.org/10.1109/ASE56229.2023.00120}
\showDOI{\tempurl}


\bibitem[Xie(2006)]%
        {DBLP:conf/ecoop/Xie06}
\bibfield{author}{\bibinfo{person}{Tao Xie}.} \bibinfo{year}{2006}\natexlab{}.
\newblock \showarticletitle{Augmenting Automatically Generated Unit-Test Suites with Regression Oracle Checking}. In \bibinfo{booktitle}{\emph{{ECOOP} 2006 - Object-Oriented Programming, 20th European Conference, Nantes, France, July 3-7, 2006, Proceedings}} \emph{(\bibinfo{series}{Lecture Notes in Computer Science}, Vol.~\bibinfo{volume}{4067})}, \bibfield{editor}{\bibinfo{person}{Dave Thomas}} (Ed.). \bibinfo{publisher}{Springer}, \bibinfo{pages}{380--403}.
\newblock
\urldef\tempurl%
\url{https://doi.org/10.1007/11785477\_23}
\showDOI{\tempurl}


\bibitem[Yang et~al\mbox{.}(2023)]%
        {DBLP:conf/issta/YangCFJS23}
\bibfield{author}{\bibinfo{person}{Chen Yang}, \bibinfo{person}{Junjie Chen}, \bibinfo{person}{Xingyu Fan}, \bibinfo{person}{Jiajun Jiang}, {and} \bibinfo{person}{Jun Sun}.} \bibinfo{year}{2023}\natexlab{}.
\newblock \showarticletitle{Silent Compiler Bug De-duplication via Three-Dimensional Analysis}. In \bibinfo{booktitle}{\emph{Proceedings of the 32nd {ACM} {SIGSOFT} International Symposium on Software Testing and Analysis, {ISSTA} 2023, Seattle, WA, USA, July 17-21, 2023}}, \bibfield{editor}{\bibinfo{person}{Ren{\'{e}} Just} {and} \bibinfo{person}{Gordon Fraser}} (Eds.). \bibinfo{publisher}{{ACM}}, \bibinfo{pages}{677--689}.
\newblock
\urldef\tempurl%
\url{https://doi.org/10.1145/3597926.3598087}
\showDOI{\tempurl}


\bibitem[Yang et~al\mbox{.}(2011)]%
        {DBLP:conf/pldi/YangCER11}
\bibfield{author}{\bibinfo{person}{Xuejun Yang}, \bibinfo{person}{Yang Chen}, \bibinfo{person}{Eric Eide}, {and} \bibinfo{person}{John Regehr}.} \bibinfo{year}{2011}\natexlab{}.
\newblock \showarticletitle{Finding and understanding bugs in {C} compilers}. In \bibinfo{booktitle}{\emph{Proceedings of the 32nd {ACM} {SIGPLAN} Conference on Programming Language Design and Implementation, {PLDI} 2011, San Jose, CA, USA, June 4-8, 2011}}, \bibfield{editor}{\bibinfo{person}{Mary~W. Hall} {and} \bibinfo{person}{David~A. Padua}} (Eds.). \bibinfo{publisher}{{ACM}}, \bibinfo{pages}{283--294}.
\newblock
\urldef\tempurl%
\url{https://doi.org/10.1145/1993498.1993532}
\showDOI{\tempurl}


\end{thebibliography}


\end{document}